\documentclass[amsmath,amssymb,floatfix,preprint,superscriptaddress]{article}
\usepackage{geometry}      
\usepackage{graphicx}
\usepackage{enumerate}
\usepackage[mathscr]{euscript}
\usepackage{amsthm}
\usepackage{amssymb}
\usepackage[english]{babel}
\usepackage{hyperref}
\usepackage[T1]{fontenc}
\usepackage[cp1250]{inputenc}
\usepackage{subfig}

\usepackage[backend=bibtex,
citestyle=numeric, 
bibstyle=numeric, 
firstinits=true, 
natbib=true,
maxcitenames=2,
maxbibnames=7,
hyperref=true,
backref=false,
arxiv=pdf]{biblatex} 
\usepackage{caption}
\usepackage{epstopdf}
\usepackage{multicol}        
\usepackage[bottom]{footmisc}
\usepackage{pifont}
\usepackage{algorithmicx}
\usepackage{algpseudocode}
\usepackage{algcompatible}
\usepackage{algorithm}
\usepackage[makeroom]{cancel}
\usepackage{float}
\usepackage{latexsym}
\usepackage{array}
\usepackage{colortbl}
\usepackage{fancybox}
\usepackage{tabulary}
\usepackage{multirow}
\usepackage{multicol}
\usepackage{tikz}
\usepackage{hhline}
\usepackage{booktabs}
\usepackage{hyphenat}
\usepackage{enumitem}
\usepackage{amsmath, amssymb}
\usepackage{dcolumn}
\usepackage{color}
\usepackage{xcolor,colortbl, arydshln}
\usepackage{framed}
\usepackage{tabu}

\allowdisplaybreaks

\hypersetup{pdfauthor={Radivojevi\'c and Akhmatskaya},pdftitle={MMHMC}}

\begin{document}

\title{Modified Hamiltonian Monte Carlo for Bayesian Inference
}

\author{ 
	Tijana Radivojevi\'c$^{1,2,3}$, 
       Elena Akhmatskaya$^{1,4}$
         \hspace{0.3cm}\\
      {\small $^1$BCAM - Basque Center for Applied Mathematics} \\
      {\small $^2$Biological Systems and Engineering Division, Lawrence Berkeley National Laboratory} \\
      {\small $^3$DOE Agile Biofoundry } \\
      {\small $^4$IKERBASQUE, Basque Foundation for Science }
   }

\maketitle

{\small 
	{\hspace{6cm} \textbf{Abstract}}
	
The Hamiltonian Monte Carlo (HMC) method has been recognized as a powerful sampling tool in computational statistics. 
We show that performance of HMC can be significantly improved by incorporating importance sampling and an irreversible part of the dynamics into a chain. This is achieved by replacing Hamiltonians in the Metropolis test with modified Hamiltonians, and a complete momentum update with a partial momentum refreshment. 
We call the resulting generalized HMC importance sampler---Mix \& Match Hamiltonian Monte Carlo (MMHMC). 
The method is irreversible by construction and further  
benefits from (i) the efficient algorithms for computation of modified Hamiltonians; (ii) the implicit momentum update procedure and (iii) the multi-stage splitting integrators specially derived for the methods sampling with modified Hamiltonians.
MMHMC has been implemented, tested on the popular statistical models and compared in sampling efficiency with HMC, Riemann Manifold Hamiltonian Monte Carlo, Generalized Hybrid Monte Carlo, Generalized Shadow Hybrid Monte Carlo, Metropolis Adjusted Langevin Algorithm and Random Walk Metropolis-Hastings. 
To make a fair comparison, we propose a metric that accounts for correlations among samples and weights, and can be readily used for all methods which generate such samples.
The experiments reveal the superiority of MMHMC over popular sampling techniques, especially in solving high dimensional problems.
}

\medskip
\noindent\textit{Keywords:} Bayesian inference, Markov chain Monte Carlo, Hamiltonian Monte Carlo, importance sampling, modified Hamiltonians

\section{Introduction}\label{Sec:Intro}

Despite the complementary nature, Hamiltonian dynamics and Metropolis Monte Carlo had never been considered jointly until the \emph{Hybrid Monte Carlo} method was formulated  in the seminal paper by \citet{DKPR87}. It was originally applied to lattice field theory simulations and remained unknown for statistical applications till 1994, when R.\ Neal used the method in neural network models \citep{Neal94}. Since then, the common name in statistical applications is \emph{Hamiltonian Monte Carlo} (HMC). 
The practitioners-friendly guides to HMC were provided by \citet{Neal10} and \citet{Betancourt:2017}, while comprehensive geometrical foundations were set by \citet{BBLG16}. The conditions under which HMC is geometrically ergodic are also established \citep{LBBG16}.

\sloppy Nowadays, HMC is used in a wide range of applications---from molecular simulations to statistical problems appearing in many fields, such as ecology, cosmology, social sciences, biology, pharmacometrics, biomedicine, engineering, business. 
The software packages \textsf{Stan} \citep{Stan} and \textsf{PyMC3} \citep{salvatier2016probabilistic} have contributed to the increased popularity of the method through the implementation of HMC based sampling in a probabilistic modeling language to help statisticians writing their models in familiar notations.

For a range of problems in computational statistics the HMC method has proved to be a successful and valuable technique. The efficient use of gradient information of the posterior distribution allows it to overcome the random walk behavior typical of the Metropolis-Hastings  Monte Carlo method.

On the other hand, the performance of HMC deteriorates, 
in terms of acceptance rates, with respect to the system's size and step size, due to errors introduced by numerical approximations \cite{IH04}. 
Many rejections induce high correlations between samples and reduce the efficiency of the estimator.
Thus, in systems with a large number of parameters, or latent parameters, or when the observations data set  is very big, efficient sampling might require a substantial number of evaluations of the posterior distribution and its gradient. This may be computationally too demanding for HMC.
In order to maintain the acceptance rate for larger systems at a high level, one could decrease a step size or use a higher order integrator, but both solutions are usually impractical for complex systems. 

Ideally, one would like to have a sampling method that maintains high acceptance rates, achieves fast convergence, demonstrates good sampling efficiency and requires  modest computational and tuning efforts. 

\begin{figure*}
	\begin{center}
		\includegraphics[width=0.9\columnwidth]{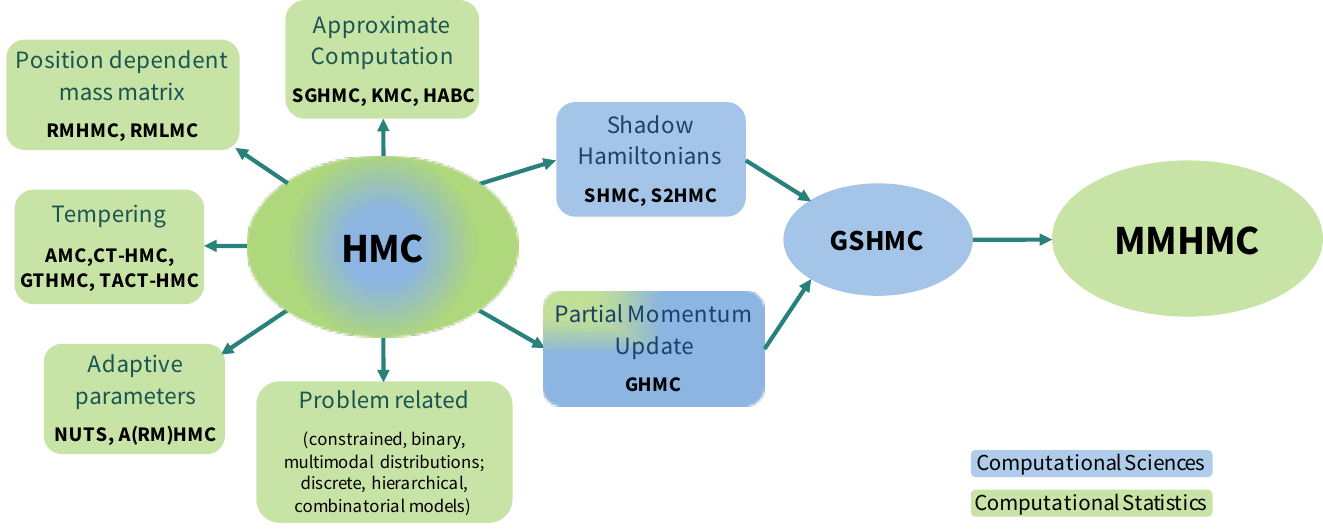}
		\caption{Evolution and relationships between some variants of the HMC methods.}
		\label{Fig:HMC_evolution}
	\end{center}
\end{figure*}

To achieve some of those goals, several modifications of the HMC method have been recently developed in  \textbf{\emph{computational statistics}} (see Figure \ref{Fig:HMC_evolution}).

It is worth of mentioning here the methods employing a 
\emph{position dependent `mass' matrix} \citep{GC11,Betancourt13a,Lan:2012}, 
\emph{adaptive} HMC \citep{HoffmanGelman14,Betancourt13,WdF12,WMdF13}, 
HMC with the 
\emph{approximated gradients}  \citep{CFG14,SSLS15,Zhang:2015,Zhang:2015a,Zhang:2017b,Zou:2018}, 
\emph{tempered} HMC \citep{vdMPW14,Betancourt:2015,Graham:2017,Nishimura:2017,Luo:2017},
HMC with \emph{alternative kinetic energy} \citep{Zhang:2016,Lu:2017,Livingstone:2017},
\emph{problem related} HMC \citep{Betancourt:2010,brubaker2012family,LZS14,PP13,LSS14,BG13,ZS14,zhang2012continuous,Afshar:2015,Nishimura:2018,Dinh:2017,Yi:2017,Kleppe:2018}, 
\emph{enhanced sampling} HMC  \citep{SDC12,SDMdW14,CSS14,Fu:2016,NishimuraDunson15,Zhang:2017,Tripuraneni:2017,Levy:2018}, and 
\emph{special cases} of HMC, such as, Metropolis Adjusted Langevin Algorithm \citep{Kennedy:1990}.

Among the modifications introduced in \textbf{\emph{computational physical sciences}}, the most important ones are  \emph{partial momentum update} and sampling with \emph{modified energies} (Figure \ref{Fig:HMC_evolution}). 

The partial momentum update (in contrast to the complete momentum update in HMC) was introduced by \citet{Horowitz91} within Generalized guided Monte Carlo, also known as the second order Langevin Monte Carlo (L2MC). The purpose of this method was to retain more dynamical information on a simulated system.

\citet{Kennedy01} formalized this idea in the Generalized Hybrid Monte Carlo (GHMC) method. GHMC is defined as the concatenation of two steps: Molecular Dynamics Monte Carlo and Partial Momentum Update.

Applications of the GHMC method to date include mainly molecular simulations. Behavior of non-special cases of GHMC are not well studied in statistical computations, with only a few exceptions (e.g.\ \cite{Sohl-Dickstein12,SDMdW14}). 

The idea of using the modified (shadow) Hamiltonian for sampling in HMC was suggested by \citet{IH04}. The performance of the resulting Shadow Hybrid Monte Carlo (SHMC) is limited by the need for a finely tuned parameter introduced for {\color{black}
	controlling the difference in the true and modified Hamiltonians} and for the evaluation of a non-separable modified Hamiltonian.
The SHMC was modified by \citet{SHSI09} through replacing a non-separable shadow Hamiltonian with the separable 4th order shadow Hamiltonian to result in Separable Shadow Hybrid Monte Carlo (S2HMC).

The first method to incorporate both, the partial momentum update and sampling with respect to a modified density, was introduced by \citet{AR06} and called Targeted Shadow Hybrid Monte Carlo (TSHMC). However, the Generalized Shadow Hybrid Monte Carlo (GSHMC) method formulated by  \citet{Akhmatskaya08} appears the most efficient  (\cite{Wee08,Akhmatskaya:2009,ARN11,Akhmatskaya:2012}) among the methods, which sample with modified Hamiltonians and are often referred to as Modified Hamiltonian Monte Carlo (MHMC) methods \cite{Akhmatskaya:2017}.

The potential {advantage} of  GSHMC compared to HMC is the enhanced sampling resulting from: (i) higher acceptance rates, achieved due to better conservation of modified Hamiltonians than Hamiltonians by symplectic integrators; (ii) an access to second-order information about the target distribution; (iii) an additional tunable parameter for improving performance; and (iv) irreversibility. The latter property of the method has never been mentioned whatsoever. Nevertheless, there is a great evidence that irreversible samplers may provide better mixing properties than their reversible counterparts do \cite{Ottobre:2016}.
On the other hand, potential {disadvantages} of GSHMC include an extra parameter to tune and the computational overhead due to repetitive evaluations of modified Hamiltonians and a momentum update Metropolis function. 

The efficiency of GSHMC method in solving statistical inference problems has never been investigated although its applicability has been recognized \cite{Akhmatskaya:2012}. 

In this paper, we present the Mix \& Match Hamiltonian Monte Carlo (MMHMC) method which is based on the GSHMC method but modified, enriched with the new features and adapted specially to computational statistics. The modifications of GSHMC that led to the MMHMC method include:

\begin{itemize}
	\item a new formulation of the importance sampling distribution relying on the modified Hamiltonians for splitting integrating schemes;

	\item numerical integration of Hamiltonian dynamics using novel multi-stage integrators, specifically derived for improving conservation of modified Hamiltonians in the MHMC methods;
	
	\item an incorporation of momentum updates in the Metropolis test for a less frequent calculation of derivatives.
	
\end{itemize}

Additionally, we propose a new metric for measuring sampling efficiency of methods which generate samples that are both correlated and weighted. 

We implemented MMHMC in our software package \textsf{HaiCS}, which also offers implementation of several other HMC based samplers as well as a range of popular statistical models.

The paper is structured as follows. We start with the summary of the Hamiltonian Monte Carlo method in Section \ref{SubSec:HMC}. The MMHMC method is formulated in Section \ref{SubSec:MMHMC} and its essential features are reviewed in Section \ref{SubSec:Features}.  
The ways of tuning and measuring performance of MMHMC are discussed in Section \ref{SubSec:Tuning}. 
The expected performance of the method is analyzed in Section \ref{Sec:PerfMMHMC}.
The details of software implementation and testing procedure as well as the test results obtained for MMHMC and compared with various popular sampling techniques are presented in Section \ref{Sec:Experiments}. The conclusions are summarized in Section \ref{Sec:Conclusions}.

\section{Mix \& Match Hamiltonian Monte Carlo (MMHMC)}\label{Sec:Formulation}
Before introducing and analyzing Mix \& Match Hamiltonian Monte Carlo we briefly revise the Hamiltonian Monte Carlo method. 

\subsection{Hamiltonian Monte Carlo: Essentials}\label{SubSec:HMC}
The purpose of HMC is to sample a random variable (r.\ v.) $\boldsymbol{\theta} \in \mathbb{R}^D$ with the distribution $\pi(\boldsymbol{\theta})$, or to estimate integrals of the form
\begin{equation}\label{Integral}
I=\int f(\boldsymbol{\theta})\pi(\boldsymbol{\theta})\mathrm d \boldsymbol{\theta}.
\end{equation}
We use the same notation $\pi$ for the probability density function (p.d.f.), which can be written as
\begin{equation*}
\pi(\boldsymbol{\theta})=\frac{1}{Z}\exp(-U(\boldsymbol{\theta})),
\end{equation*}
where the variable $\boldsymbol{\theta}$ corresponds to the position vector, $U(\boldsymbol{\theta})$ to the potential function of a Hamiltonian system and $Z$ is the normalizing constant such that $\pi(\boldsymbol{\theta})$ integrates to one.
In Bayesian framework, the target distribution $\pi(\boldsymbol{\theta})$ is the posterior distribution $\pi(\boldsymbol{\theta}|\mathbf y)$ of unknown parameters given  data $\mathbf y=\{y_1,\dots,y_{K}\ \}$, $K$ is the size of the data, and the potential function can be defined as
\begin{equation*}\label{eq:potential_func}
U(\boldsymbol{\theta})=-\log L(\boldsymbol{\theta}|\mathbf y) -\log p(\boldsymbol{\theta}),
\end{equation*} 
for the likelihood function $ L(\boldsymbol{\theta}|\mathbf y)$ and prior p.d.f.\ $p(\boldsymbol{\theta})$ of model parameters.

The auxiliary momentum variable $\mathbf p \in \mathbb{R}^D$, conjugate to and independent of the vector $\boldsymbol{\theta}$ is typically drawn from a normal distribution
\begin{equation}\label{mom_pdf}
\mathbf p \sim \mathcal{N}(0,M),
\end{equation}
with a covariance matrix $M$, which is positive definite and often diagonal.
The Hamiltonian function can be defined in terms of the target p.d.f.\ as the sum of the potential function $U(\boldsymbol{\theta})$ and the kinetic function $K( \mathbf p)$

\begin{equation}\label{fullHam}
\begin{aligned}
H(\boldsymbol{\theta}, \mathbf p)&=U(\boldsymbol{\theta}) +K( \mathbf p)\\
&= U(\boldsymbol{\theta}) +\frac{1}{2} \mathbf p^T M^{-1} \mathbf p+\frac{1}{2}\log\left((2\pi)^D|M|\right).
\end{aligned}
\end{equation}
The joint p.d.f.\ is then
\begin{equation}\label{joint_pdf}
\begin{aligned}
\pi(\boldsymbol{\theta}, \mathbf p) &=\frac{1}{Z} \exp(- H(\boldsymbol{\theta}, \mathbf p))\\
&=\frac{(2\pi)^\frac{D}{2}|M|}{Z}\exp(- U(\boldsymbol{\theta})) \exp(- \frac{1}{2} \mathbf p^T M^{-1} \mathbf p).
\end{aligned}
\end{equation}
By simulating a Markov chain with the invariant distribution \eqref{joint_pdf}  and marginalizing out  momentum variables, one recovers the target distribution $\pi(\boldsymbol{\theta})$. The integral \eqref{Integral} can then be estimated using $N$ simulated samples as
$$\hat{I}=\frac{1}{N}\sum_{n=1}^N f(\boldsymbol{\theta}^n).$$

HMC samples from $\pi(\boldsymbol{\theta}, \mathbf p)$ by alternating a step for a momentum refreshment and a step for a joint, position and momentum, update, for each Monte Carlo iteration. In the first step, momentum is replaced by a new draw from the normal distribution \eqref{mom_pdf}. In the second step, a proposal for the new state $(\boldsymbol{\theta}', \mathbf p')$ is generated by integrating Hamiltonian dynamics  
\begin{equation}\label{HamDyn}
\frac{\mathrm d \boldsymbol{\theta}}{\mathrm d t} = M^{-1} \mathbf p, \; \; 
\frac{\mathrm d \mathbf p}{\mathrm d t} = -U_{\boldsymbol{\theta}}(\boldsymbol{\theta})
\end{equation}
for $L$ steps  using a symplectic  integrator $\Psi_{h}$ with a step size $h$. 
Due to the numerical approximation of integration, Hamiltonian function, and thus the density \eqref{joint_pdf}, are not preserved.
In order to restore this property, which ensures invariance of the target density, an accept-reject step is added through a Metropolis criterion. 
The acceptance probability has a simple form
$$\alpha=\min\left\{1,\exp\left(H(\boldsymbol{\theta}, \mathbf p)-H(\boldsymbol{\theta}', \mathbf p')\right)\right\},$$
which, due to the preservation of volume, does not include potentially difficult to compute Jacobians of the mapping. As in any Markov chain Monte Carlo (MCMC) method, in case of a rejection, the current state is stored as a new sample. Once next sample is obtained, momentum is replaced by a new draw, so Hamiltonians have different values for consecutive samples. This means that samples are drawn along different level sets of Hamiltonians, which actually makes HMC an efficient sampler.

For a constant matrix $M$, the last term in the Hamiltonian \eqref{fullHam} is a constant that cancels out in the Metropolis test. Therefore, the Hamiltonian can be defined as
\begin{equation}\label{eq:Ham}
H(\boldsymbol{\theta}, \mathbf p)=U(\boldsymbol{\theta}) + \frac{1}{2} \mathbf p^T M^{-1}\mathbf p.
\end{equation}

The algorithmic summary of the HMC method is given in Appendix \ref{Sec:Algorithm}.

\subsection{Formulation of MMHMC}\label{SubSec:MMHMC}

\begin{figure}[h!]
	\begin{center}
		\includegraphics[width=0.8\columnwidth ]{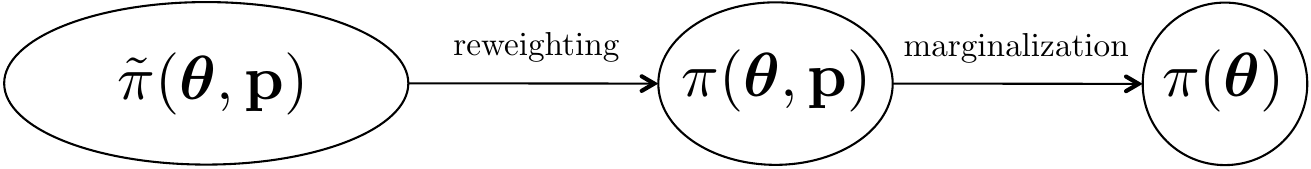}
		\caption{MMHMC indirect sampling of the target distribution.}
		\label{MMHMCsampling}
	\end{center}
\end{figure}

As HMC, the MMHMC method aims at sampling unknown parameters $\boldsymbol{\theta} \in \mathbb R^D$ with the distribution (known up to a normalizing constant)
$$\pi(\boldsymbol{\theta} )\propto \exp(-U(\boldsymbol{\theta} )).$$
However, this is achieved indirectly, as shown in Figure \ref{MMHMCsampling}.  

More precisely, MMHMC performs HMC importance sampling on the joint state space of positions and momenta $(\boldsymbol{\theta},\mathbf p)$ with respect to the modified density $\tilde\pi$. The target distribution on the joint state space $\pi(\boldsymbol{\theta},\mathbf p)\propto \exp(-H(\boldsymbol{\theta},\mathbf p))$, with respect to the true Hamiltonian $H$, is recovered through importance reweighting and finally, the desired distribution $\pi(\boldsymbol{\theta})$ is retrieved by marginalizing momenta variables. The MMHMC algorithm consists of three major steps: (1) Hamiltonian Dynamics Monte Carlo (HDMC) step to generate the next state, (2) Partial Momentum Monte Carlo (PMMC) step to refresh a momentum for each state, and (3)  importance reweighting to recover the target distribution. The essential constituents of the algorithm are explained below.  

\subsubsection{Importance Distribution}\label{Sec:ImportDistr}

The importance distribution in MMHMC ought to satisfy two principal requirements. First, it should lead to more favourable values of the acceptance probability than may be achieved in the HMC algorithm. Second, the target density and the importance density have to be close to maintain a low variability among weights, essential for efficient sampling. The so called modified Hamiltonian is a promising candidate for serving these purposes.

Given Hamiltonian dynamics with Hamiltonian function $H$ \eqref{eq:Ham} and a symplectic integrator with an integration step size $h$ for solving consequent ODE equations \eqref{HamDyn}, the corresponding modified equations are guaranteed to be Hamiltonian and the modified Hamiltonian can be
determined as \cite{HLW06}
\begin{equation}\label{eq:DefModHam}
\tilde{H_h}= H + h H_2+ h^2 H_3+\cdots.
\end{equation}
In contrast to the Hamiltonian, the modified Hamiltonian is exactly preserved along the computed trajectory by symplectic integrators \cite{LR05}. For an integrator of order $m$  (\(m \geq 2\)), 
\[\tilde{H_h} = H+\mathcal{O}(h^m).\] 
For the $k$-truncation of \(\tilde H_h\) (\(k>m\)) 
defined as 
\begin{equation}\label{eq:TruncatedModHam}
\tilde{H}_h^{[k]}=H+...+h^mH_{m+1}+\dots+h^{k-1}H_k, 
\end{equation}
one obtains
\begin{equation}\label{eq:TruncatedBound}
\tilde{H}_h^{[k]} = H+\mathcal{O}(h^k),
\end{equation}
and hence, a symplectic method preserves the $k$-truncated modified Hamiltonian up to order $h^k$.
The expectation of the increment of $\tilde{H}_h^{[k]}$ in an integration leg satisfies
\begin{equation}\label{eq:ExpectationModifiedHamiltonian}
\mathbb{E}_{\tilde{\pi}}[\Delta \tilde H_h^{[k]}] = \mathcal O\left(Dh^{2k}\right),
\end{equation}
with $D$ being the dimension, while for the Hamiltonian it is
\begin{equation}\label{eq:ExpectationHamiltonian}
\mathbb{E}_{\pi}[\Delta H] = \mathcal{O}\left(Dh^{2m}\right)
\end{equation}
\cite{BPRSSS13}, and therefore the MMHMC algorithm may benefit from high acceptance rates due to better conservation of \(\tilde{H}^{[k]}\).

The importance canonical density in MMHMC is then chosen as 
\begin{equation}\label{modDistrMMHMC}
\tilde\pi(\boldsymbol{\theta},\mathbf p)\propto \exp(-\tilde {H}_h^{[k]}(\boldsymbol{\theta},\mathbf p)).
\end{equation} 

For simplicity, we drop the subscript $h$ and superscript $[k]$ in $\tilde {H}_h^{[k]}$ assuming an arbitrary choice of a truncation order.
We shall return to the issue in the discussion of the specific formulations of the modified Hamiltonians associated with particular choices of a numerical integrator.

We notice that randomization of a step size commonly applied in HMC simulations is not compatible with the proposed importance distribution \eqref{modDistrMMHMC}. On the one hand, randomization of a step size implies that a general modified equation does not exist, and thus the modified Hamiltonian can be constructed locally only, hence, the importance density has to be modified accordingly. On the other hand, randomization of a step size inevitably leads to the increased variability of weights, meaning the ultimate performance degradation of the importance sampling algorithm. Therefore, in MMHMC, the priority is given to a fixed step size. The advantages of this strategy are demonstrated in Section \ref{Sec:PerfMMHMC}.

\subsubsection{Hamiltonian Dynamics Monte Carlo (HDMC)} 

At every MC iteration, a proposal $(\boldsymbol{\theta}', \mathbf p')$ is generated by simulating Hamiltonian dynamics \eqref{HamDyn} using a symplectic and reversible numerical integrator $\psi_h$ with a step size $h$, and is accepted with the Metropolis criterion corresponding to the modified distribution \eqref{modDistrMMHMC} as
\begin{equation}\label{HDMetropolis}
(\boldsymbol{\theta}^{new}, \mathbf p^{new})=\left\{
\begin{array}{l l}
(\boldsymbol{\theta}', \mathbf p') & \quad  \text{ with probability} \quad \alpha \\
\mathcal F(\boldsymbol{\theta}, \mathbf p) & \quad \text{ otherwise,}
\end{array} 
\right.
\end{equation}
where $\alpha=\min \left\{1, \exp(-\Delta\tilde H) \right\}$ and $\mathcal F(\boldsymbol{\theta}, \mathbf p)$ flips the momentum in the case of rejection, i.e.  $\mathcal F(\boldsymbol{\theta}, \mathbf p)=(\boldsymbol{\theta}, -\mathbf p)$, and $\Delta\tilde H=\tilde H(\boldsymbol{\theta}', \mathbf p') -\tilde H(\boldsymbol{\theta}, \mathbf p)$.

An integrator $\psi_h$ can be chosen arbitrarily from the class of symplectic and reversible integration schemes, though computationally efficient and accurate $\psi_h$ are highly desirable for achieving the top performance of MMHMC. While GSHMC was formulated with the leapfrog integrator in mind, in MMHMC we employ multi-stage splitting schemes, proposed by \citet{Radivojevic:2018}. 
More specifically, we consider numerical schemes belonging to the two-stage  
\begin{equation}\label{2S}
\psi_h=\varphi^B_{bh}\circ\varphi^A_{\frac{h}{2}}\circ \varphi^B_{(1-2b)h}\circ\varphi^A_{\frac{h}{2}}\circ \varphi^B_{bh}
\end{equation}
and three-stage
\begin{equation}\label{3S}
\psi_h=\varphi^B_{bh}\circ\varphi^A_{a h}\circ \varphi^B_{(\frac{1}{2}-b)h}\circ\varphi^A_{(1-2a)h}\circ \varphi^B_{(\frac{1}{2}-b)h}\circ\varphi^A_{a h}\circ \varphi^B_{bh}
\end{equation}
families of splitting methods, which can offer better conservation properties than Verlet / leapfrog \cite{BCSS14}.
Here, the exact flows $\varphi^A_{h}$ and $\varphi^B_{h}$ are solutions to the split systems
\begin{equation*}\label{HamA}
A: \frac{\mathrm d \boldsymbol{\theta}}{\mathrm d t} = 0, \; \;
\frac{\mathrm d \mathbf p}{\mathrm d t} = -U_{\boldsymbol{\theta}}(\boldsymbol{\theta}),
\end{equation*}
and
\begin{equation*}\label{HamB}
B: \frac{\mathrm d \boldsymbol{\theta}}{\mathrm d t} = M^{-1} \mathbf p, \; \;
\frac{\mathrm d \mathbf p}{\mathrm d t} = 0,
\end{equation*}
respectively, corresponding to the Hamiltonian \eqref{eq:Ham}, and $a$, $b$ are parameters of an integrator $\psi_h$, which will be discussed later. 

\subsubsection{Partial Momentum Monte Carlo (PMMC)} 

Whereas in HMC momentum is completely reset at each MC step before numerical integration, MMHMC relies on the partial refreshment of momentum. The idea behind the partial momentum update is to suppress the random walk behaviour arising from the complete, and hence independent from the current momentum, update. 
The PMMC can be performed in two steps. 

First, for the current momentum $\mathbf {p}$ and a noise vector $\mathbf {u} \sim \mathcal N(0,M)$ a proposal for the new momentum $\mathbf p^*$ is generated from the mapping $\mathcal{R}:(\boldsymbol{\theta}, \mathbf p, \mathbf u) \mapsto(\boldsymbol{\theta}, \mathbf p^*, \mathbf u^*)$ such that
\begin{equation}\label{eq:PMMCorig}
\mathcal{R}(\boldsymbol{\theta}, \mathbf p, \mathbf u)=(\boldsymbol{\theta}, \sqrt{1-\varphi}\mathbf{p} + \sqrt{\varphi}\mathbf {u}, -\sqrt{\varphi}\mathbf{p} + \sqrt{1-\varphi}\mathbf {u}).
\end{equation}
Here, parameter $\varphi \in (0,1]$ controls the amount of noise introduced in every MC iteration. 

Then, to secure sampling from the modified density \eqref{modDistrMMHMC}, the proposal is 
accepted according to the extended modified distribution
\begin{equation}\label{extended_target_prob}
\hat{\pi}\propto \exp(-\hat{H}),
\end{equation}
with the extended Hamiltonian $\hat H$ defined as
\begin{equation}\label{extendedHam}
\hat{H}(\boldsymbol{\theta},\mathbf p,\mathbf u)=\tilde{H}(\boldsymbol{\theta},\mathbf p)+{\frac{1}{2}}\mathbf u^{\intercal}M^{-1}\mathbf u.
\end{equation} 
Therefore, a new momentum can be determined as
\begin{equation}\label{newPMMC_Metropolis}
\bar{\mathbf p}=\left\{\begin{array}{l l}
\mathbf p^*=\sqrt{1-{\varphi}}\mathbf p+\sqrt{{\varphi}}\mathbf u  & \mbox{ with probability } \\
& \hspace{0.5cm} \mathcal{P}= \min\{1,\exp(-\Delta\hat{H})\}\\
\mathbf p & \mbox{ otherwise,}
\end{array}\right.
\end{equation}
where $\Delta \hat H=\hat H(\boldsymbol{\theta}, \mathbf p^*, \mathbf u^*) -\hat H(\boldsymbol{\theta}, \mathbf p, \mathbf u)$. 

Formulated in such a way, the PMMC step introduces two extra evaluations of the modified Hamiltonian within the Metropolis test and thus a computational overhead. To reduce the overhead, we incorporated a momentum proposal in the Metropolis test and derived the computationally tractable expressions for $\Delta \hat H$, for the particular choices of modified Hamiltonian, which we recommend to use in MMHMC. The details are provided in Section \ref{SubSec:Features} and Appendix \ref{App:ModPMMC}.

\subsubsection{Reweighting} 

After $N$ iterations of the MMHMC algorithm, 
{reweighting} is required in order to estimate the integral \eqref{Integral}. By
making use of the standard technique for importance samplers, the integral is rewritten as
\begin{equation*}\label{expect_MMHMC}
\begin{aligned}
I=\mathbb E_{\pi}[f]&=\int f(\boldsymbol{\theta})\pi(\boldsymbol{\theta},\mathbf p)\mathrm d \boldsymbol{\theta}\mathrm d\mathbf p\\
&=\int f(\boldsymbol{\theta})\frac{\pi(\boldsymbol{\theta},\mathbf p)}{\tilde{\pi}(\boldsymbol{\theta},\mathbf p)}\tilde{\pi}(\boldsymbol{\theta},\mathbf p)\mathrm d \boldsymbol{\theta}\mathrm d\mathbf p \\
&= \int f(\boldsymbol{\theta})w(\boldsymbol{\theta},\mathbf p)\tilde{\pi}(\boldsymbol{\theta},\mathbf p)\mathrm d \boldsymbol{\theta}\mathrm d\mathbf p =\mathbb E_{\tilde{\pi}}[fw],
\end{aligned}
\end{equation*}
where $\tilde{\pi}(\boldsymbol{\theta},\mathbf p)$ is the importance distribution \eqref{modDistrMMHMC} and $w(\boldsymbol{\theta},\mathbf p)$ the importance weight function.
Therefore, the integral can be approximated by a self-normalized estimator as 
\begin{equation}\label{MMHMCestimator}
\begin{aligned}
\hat I&=\frac{\sum_{n=1}^N f(\boldsymbol{\theta}^n)w_n}{\sum_{n=1}^N w_n}, \; \; \; \\
\end{aligned}
\end{equation}
\begin{equation}\label{Eq:Weights}
\begin{aligned}
w_n&=\exp(\tilde{H}(\boldsymbol{\theta}^n,\mathbf p^n)-H(\boldsymbol{\theta}^n,\mathbf p^n)),	
\end{aligned}
\end{equation}
where $\{(\boldsymbol{\theta}^n,\mathbf p^n)\}_{n=1}^N$ is drawn from $\tilde{\pi}$, and $w_n$ are the corresponding weights.  

Performance of importance sampling methods strongly depends on the discrepancy between the target and importance sampling distributions, and thus on weights. Bounded weights imply a bounded variance of an estimator. The choice of the importance distribution \eqref{modDistrMMHMC} in MMHMC along with \eqref{eq:TruncatedBound} guarantee that the MMHMC weights are bounded and thus the reduction in efficiency of the estimator \eqref{MMHMCestimator},  introduced due to importance sampling, is minor in the case of the MMHMC method.

\subsection{Features of MMHMC}\label{SubSec:Features}
In this section we discuss in more detail the specific features of the MMHMC method.  
The main algorithmic differences between HMC and MMHMC  are listed in Table \ref{MMHMCvsHMC} and full algorithmic summary of MMHMC is provided in Appendix \ref{Sec:Algorithm}. 

\begin{table}[ht]
	\caption{Algorithmic differences between HMC and MMHMC.}
	\begin{center}
		\begin{tabular}{c c c c }
			\toprule
			& HMC&& MMHMC\\ 
			\midrule
			Momentum update &complete &&partial \\ 
			Momentum Metropolis test &\ding{55}  & & \ding{51} \\
			Metropolis test& $H$& &$\tilde{H}$\\
			Momentum flips & \ding{55} &&  \ding{51}  \\  
			Re-weighting & \ding{55} &&  \ding{51}  \\ 
			Reversibility & \ding{51}  && \ding{55} \\
			\bottomrule
		\end{tabular}
		\label{MMHMCvsHMC}
	\end{center}
\end{table}

\subsubsection{Irreversibility}

Until recently, the significant attention in the literature has been paid to the theoretical analysis of reversible Markov chains rather than the study of irreversible MCMC methods.
However, numerous latest theoretical and numerical results demonstrate the advantage of irreversible MCMC over reversible algorithms both in terms of
variance of an estimator and rates of convergence to the target distribution \cite{Neal:2004,Suwa:2012,Ohzeki:2015,Bouchard-Cote:2016,Ottobre:2016,Duncan:2016,Duncan:2017}. 
These well documented facts have induced a design of new algorithms which break the detailed balance condition (DBC)---a commonly used criterion to demonstrate the invariance of the chain.
Some recent examples of irreversible methods based on Hamiltonian dynamics can be found in papers by \citet{Ottobre:2016,OPPS14,Ma:2016}.

The core of the MMHMC algorithm consists of two steps,  PMMC and HDMC, which both leave the target distribution $\tilde{\pi}$ invariant. 
However, the resulting chain is not reversible.

Apart from being invariant with respect to the target distribution, the HDMC step satisfies the modified DBC. The proof for the GHMC method can be found elsewhere \cite{FSSS14}, and the only difference in the case of MMHMC is that the target distribution, and thus the acceptance probability, is defined with respect to the modified Hamiltonian.

As the PMMC step is specific only to MMHMC and GSHMC, we provide a direct proof of invariance of this step (Appendix \ref{App:Invariance}). Furthermore, in an analogous way to HDMC, it can be proved that PMMC satisfies the modified DBC. 
The key observation is that the proposal mapping $\mathcal R$ for momenta \eqref{eq:PMMCorig} is reversible w.r.t.\ the extended target $\hat{\pi}$, $\mathcal R^{-1}=\mathcal{\hat{F}}^{-1}\circ\mathcal R\circ\mathcal{\hat{F}}$, and the reversing mapping $\mathcal{\hat{F}}(\boldsymbol{\theta},\mathbf p,\mathbf u):=(\boldsymbol{\theta},\mathbf p, -\mathbf u)$ is an involution.

The irreversibility of MMHMC arises from an important property---a non-symmetric composition of steps satisfying  DBC does not preserve  DBC. Therefore, although both steps of MMHMC do satisfy the (modified) DBC, their compositon is not symmetric and hence, the chain generated by MMHMC is not reversible by construction.

\subsubsection{Numerical Integrators}

The detailed discussion on efficiency of various numerical integrators in the MHMC methods can be found elsewhere \cite{Radivojevic:2018}. Here we review the most promising integration schemes for the MMHMC method and provide some practical recommendations. 

The Verlet/leapfrog integrator, considered as the integrator of choice for MHMC methods until recently, still can be seen as a perfect option for MMHMC in sampling small sized problems, where comparatively long step sizes are allowed. For such problems, Verlet is expected to demonstrate the highest conservation of modified Hamiltonians due to its best stability among splitting integrators. For bigger dimensions and thus for smaller optimal step sizes, the multi-stage integrators \eqref{2S}--\eqref{3S} designed specifically for MHMC and referred to as modified splitting integrators \cite{Radivojevic:2018} should provide better conservation of modified Hamiltonian than the Verlet integrator, resulting in enhanced accuracy and sampling performance of MHMC methods. 

Modified splitting integrators are characterized by values of parameters $a$ and $b$ in \eqref{2S}--\eqref{3S} obtained through minimization of the (expected) modified Hamiltonian error introduced by integration.  
Following the ideas of \citet{McL95} and \citet{BCSS14} for improving HMC performance by minimizing (expected) energy error through the appropriate choice of parameters of the integrator, the modified splitting integrators have been derived  \citep{Radivojevic:2018} by considering
either the error in the modified Hamiltonians for splitting integrators, $\tilde{H}^{[l]}$, of order $l = 4, 6$
\begin{equation*}\label{modHamError}
\Delta=\tilde{H}^{[l]}(\Psi_{h,L}(\boldsymbol{\theta},\mathbf p))-\tilde{H}^{[l]}(\boldsymbol{\theta},\mathbf p),
\end{equation*}
to yield the integrators M-ME2 and M-ME3, 
or the expected values of such errors $\mathbb E_{\tilde{\pi}}(\Delta)$ taken with respect to the modified
canonical density $\tilde{\pi}$ \eqref{modDistrMMHMC} to give rise to the integrators M-BCSS2 and M-BCSS3. 
Here $\Psi_{h,L}(\boldsymbol{\theta},\mathbf p)$ is the  $hL$-time map of the integrator.

In Table \ref{Tab:IntSummary} we provide important characteristics of the integrators which can be recommended for the use in modified Hamiltonian Monte Carlo methods in general, and in MMHMC in particular, for a broad range of problems and methods' parameters.  
\begin{table}[!h]
	\caption{The splitting integrators for sampling with modified Hamiltonian Monte Carlo methods using 4th order modified Hamiltonians.
		Stability limit $h_{\max}$
		is presented in terms of the three-stage family \protect\cite{Radivojevic:2018}.}
	\centering
	\begin{tabular}{cccccc}
		Integrator & N.\ of stages & Coefficients & $h_{\max}$\\ \hline
		Verlet & 1 & -- & 6.000 \\ \hline
		M-BCSS2 & 2 & $b=0.238016$ & 4.144\\ \hline
		M-ME2 & 2 & $b=0.230907$ & 4.089\\ \hline
		\multirow{2}{*}{M-BCSS3} & \multirow{2}{*}{3} & $a=(1-2b)/4(1-3b)$ & \multirow{2}{*}{4.902}\\
		&& $b=0.144115$ & \\ \hline
		\multirow{2}{*}{M-ME3} & \multirow{2}{*}{3} & $a=(1-2b)/4(1-3b)$ & \multirow{2}{*}{4.887}\\
		&& $b=0.142757$ & \\ \hline
	\end{tabular}
	\label{Tab:IntSummary}
\end{table}

\subsubsection{Modified Hamiltonians}\label{Sec:ModHam}

As in any modified Hamiltonian Monte Carlo (MHMC) method, in MMHMC, the importance distribution $\tilde\pi$ is ultimately defined through a modified Hamiltonian associated with a particular numerical integrator.  In the early MHMC methods, various implementations of modified Hamiltonians for the Verlet/leapfrog integrator have been proposed and used. The idea to employ multi-stage integration splitting schemes in MHMC methods has been explored for the first time in the context of Mix \& Match Hamiltonian Monte Carlo. Nevertheless, the derived formulations of modified Hamiltonians and parameters for corresponding integration schemes can be successfully used with other MHMC methods, as it has been discussed and demonstrated by \citet{Radivojevic:2018}. 
In the following, we briefly review the formulations of the modified Hamiltonian for splitting integrators \citep{Radivojevic:2018}, which we recommend to use along with Mix \& Match Hamiltonian Monte Carlo.

Two alternative formulations of the 4th and 6th order modified Hamiltonians corresponding to the Verlet integrator and multi-stage integrators \eqref{2S}--\eqref{3S} with arbitrary coefficients, have been proposed \citep{Radivojevic:2018}. 

For problems in which analytical derivatives of the potential functions are available and inexpensive to compute, the 4th and 6th order modified Hamiltonians for splitting integrators  can be calculated as 
\begin{align}
\tilde{H}^{[4]}(\boldsymbol{\theta},\mathbf p)  &= H(\boldsymbol{\theta},\mathbf p) +h^2 c_{21} \mathbf p^T M^{-1}U_{\boldsymbol{\theta}\boldsymbol{\theta}}(\boldsymbol{\theta}) M^{-1}\mathbf p\nonumber\\
&+ h^2 c_{22} {U_{\boldsymbol{\theta}}(\boldsymbol{\theta})}^T M^{-1}U_{\boldsymbol{\theta}}(\boldsymbol{\theta}),\label{modHam4_an}   \\
\tilde{H}^{[6]}(\boldsymbol{\theta},\mathbf p) &= \tilde{H}^{[4]}(\boldsymbol{\theta},\mathbf p)\nonumber\\
&+ h^4 c_{41} U_{\boldsymbol{\theta} \boldsymbol{\theta} \boldsymbol{\theta} \boldsymbol{\theta} }(\boldsymbol{\theta} ) M^{-1}\mathbf pM^{-1}\mathbf p M^{-1}\mathbf p M^{-1}\mathbf p\nonumber\\
&+ h^4 c_{42}{U_{\boldsymbol{\theta} }(\boldsymbol{\theta} )}^T M^{-1}U_{\boldsymbol{\theta} \boldsymbol{\theta} \boldsymbol{\theta}}(\boldsymbol{\theta}) M^{-1}\mathbf pM^{-1}\mathbf p \nonumber\\
&+ h^4 c_{43}{U_{\boldsymbol{\theta} }(\boldsymbol{\theta})}^T M^{-1}U_{\boldsymbol{\theta}\boldsymbol{\theta}}(\boldsymbol{\theta}) M^{-1}U_{\boldsymbol{\theta}}(\boldsymbol{\theta})\nonumber\\
&+  h^4 c_{44} \mathbf p^T M^{-1}U_{\boldsymbol{\theta}\boldsymbol{\theta}}(\boldsymbol{\theta}) M^{-1}U_{\boldsymbol{\theta}\boldsymbol{\theta}}(\boldsymbol{\theta}) M^{-1}\mathbf p. \label{modHam6_an} 
\end{align}

If the potential function is quadratic, i.e.\ corresponding to problems of sampling from Gaussian distributions, the 6th order modified Hamiltonian \eqref{modHam6_an} simplifies to
\begin{equation}
\begin{aligned}
\tilde{H}^{[6]}(\boldsymbol{\theta},\mathbf p) &=  \tilde{H}^{[4]}(\boldsymbol{\theta},\mathbf p)\\
&+ h^4 c_{43}{U_{\boldsymbol{\theta} }(\boldsymbol{\theta})}^T M^{-1}U_{\boldsymbol{\theta}\boldsymbol{\theta}}(\boldsymbol{\theta}) M^{-1}U_{\boldsymbol{\theta}}(\boldsymbol{\theta}) \label{modHam6_an_Gauss}\\
&+  h^4 c_{44} \mathbf p^T M^{-1}U_{\boldsymbol{\theta}\boldsymbol{\theta}}(\boldsymbol{\theta}) M^{-1}U_{\boldsymbol{\theta}\boldsymbol{\theta}}(\boldsymbol{\theta}) M^{-1}\mathbf p. %
\end{aligned}
\end{equation}

The values of the coefficients $c_{ij}$ in \eqref{modHam4_an}--\eqref{modHam6_an_Gauss} for Verlet, two- and three-stage integrators are provided in Appendix \ref{App:ModHam}. 

The alternative formulations of modified Hamiltonians address to problems with a dense Hessian matrix (and higher derivatives) and mainly rely on quantities that are available during a simulation 
\citep{Radivojevic:2018}. In this case, the 4th and 6th order modified Hamiltonians, respectively, are given as
\begin{align}
\tilde{H}^{[4]}(\boldsymbol{\theta} ,\mathbf p)  &=H(\boldsymbol{\theta} ,\mathbf p)+h k_{21} \mathbf p^T M^{-1}{P_1}\nonumber \\ 
&+ h^2 k_{22} {U_{\boldsymbol{\theta}}(\boldsymbol{\theta})}^T M^{-1}U_{\boldsymbol{\theta}}(\boldsymbol{\theta} )  \label{modHam4num}\\
\tilde{H}^{[6]}(\boldsymbol{\theta} ,\mathbf p) &=\tilde{H}^{[4]}(\boldsymbol{\theta} ,\mathbf p)\nonumber \\ 
&+ h k_{41}\mathbf p^T M^{-1}P_3 + h^2 k_{42} {{U_{\boldsymbol{\theta} }}(\boldsymbol{\theta})}^TM^{-1}P_2 \label{modHam6num} \nonumber \\ 
&+ h^2 k_{43} P_1^TM^{-1}P_1\nonumber \\ 
&+  h^4 k_{44} U_{\boldsymbol{\theta} }(\boldsymbol{\theta} )^TM^{-1} U_{\boldsymbol{\theta} \boldsymbol{\theta}}(\boldsymbol{\theta} ) M^{-1}U_{\boldsymbol{\theta} }(\boldsymbol{\theta}),
\end{align}
where the coefficients $k_{ij}$ are provided in Appendix \ref{App:ModHam}. Here $P_i=\mathbf U^{(i)}\cdot h^i, i=1,2,3$, and $\mathbf U^{(i)}$ are centered finite difference approximations of time derivatives of the gradient of the potential function (see Appendix \ref{App:ModHam} for further details).

We note that the expression \eqref{modHam4num}  allows for computation of $\tilde{H}^{[4]}$ using quantities available from a simulation. Nevertheless, this is not the case for the resulting 6th order Hamiltonian. 
The last term in \eqref{modHam6num}, arising from an expansion of the Poisson bracket $\{B, B, A, A, B\}$, cannot be computed using time derivatives of available quantities and requires explicit calculation of the Hessian matrix of the potential function.
Only for the Verlet integrator does this term vanish and the resulting coefficients are
\begin{eqnarray*}\label{modHamVerletnum}
	k_{21}&=& \frac{1}{12}, \hspace{0.3cm} k_{22}=- \frac{1}{24},\nonumber \\
	k_{41}& = &-\frac{1}{720}, \hspace{0.3cm} k_{42} = \frac{1}{240}, \hspace{0.3cm} k_{43} = \frac{11}{720}, \hspace{0.3cm} k_{44} =0.
\end{eqnarray*}

Finally, we remark that the presented formulations of modified Hamiltonians \eqref{modHam4_an}--\eqref{modHam6_an_Gauss} and \eqref{modHam4num}--\eqref{modHam6num} were used to derive the computationally tractable expressions for the Metropolis function of the modified PMMC step proposed in MMHMC (see Appendix \ref{App:ModPMMC}).

\subsection{Tuning and Measuring Performance of MMHMC}\label{SubSec:Tuning}

In this section, we first discuss the impact of the parameters of the MMHMC method on its performance.  
Secondly, we present the metrics for assessing the performance, which are specifically designed for the class of MHMC methods.

\subsubsection{Choice of Parameters}\label{Sec:Parameters}

MMHMC has five tunable parameters that affect the performance of the method---the integration step size $h$,  number of integration steps $L$,  mass matrix $M$,  noise parameter $\varphi$, and  order $k$ of the modified Hamiltonian. In principle, these parameters may be chosen arbitrarily within allowed-by-the-algorithm ranges, except for some special cases when they might affect the ergodicity of the chain (e.g.\ combinations leading to a value that is a multiple of the period of a mode of the system). However, the choice of parameters may have a dramatic impact on the overall performance of MMHMC, and thus tuning free parameters in order to maximize sampling efficiency and minimize computational costs is one of the most important but challenging tasks. 

We notice that the first three parameters of MMHMC are the same as in HMC, and like for HMC, the optimal choice of these parameters in MMHMC is still an unresolved issue, though some recommendations and observations for both methods are available  \cite{Mackenzie:1989,Liu08,Neal10,HoffmanGelman14,Akhmatskaya08,Wee08}. 
Below we briefly discuss considerations and observations  which are essential for choosing free parameters in MMHMC, while not necessarily relevant to HMC. 

For example, the experiments revealed that the parameter $L$ found to be the best for HMC is not necessarily the best for MMHMC. Actually, too long values of $L$ may result in poorer overall efficiency of MMHMC at particular choices of $\varphi$, although the computational overhead is smaller with larger $L$, due to a less frequent calculation of modified Hamiltonians. 
In contrast, longer trajectories are needed for HMC to achieve its full potential, especially for larger dimensions. Intuitively, such a difference can be explained by the presence of a partial momentum update and high acceptance rates in MMHMC, which together, for small $L$, mimic as long or even longer, but more variative than in the case of large $L$, trajectory. On the contrary, a complete momentum update and short trajectories in HMC may initiate too frequent switches to not necessarily preferable directions.   

We have to stress that the choice of a step size $h$ critically affects the accuracy and sampling efficiency of MMHMC not only through its influence on acceptance rates (like in HMC) but also on importance weights (see Section \ref{Sec:PerfMMHMC}). Indeed, the reduction in efficiency due to use of importance sampling is expected to be negligible for small values of $h$. The reason is a choice of the importance density $\tilde\pi$ in MMHMC, which stays closer to the true density $\pi$ when $h$ tends to 0. The larger values of step size may lead to a high variability in the importance weights and thus to a performance degradation. As a result, given a sampling problem, the best performance of MMHMC (often superior to the one accessible with HMC) may be achieved at step sizes smaller than the optimal ones for HMC.   

Similarly to conventional HMC, the current implementation of MMHMC uses the identity mass matrix and offers different randomization schedules for a number of integration steps. In addition, a randomization of a noise parameter is provided in the algorithm. However, in contrast to HMC, in MMHMC a step size stays fixed on the reasons explained in Section \ref{Sec:ImportDistr}.  

The parameters $\varphi$ and $k$ are specific to MMHMC and are not used in HMC. 

\paragraph{Noise parameter $\varphi$.} Too small values of $\varphi$ may reduce sampling efficiency by producing almost deterministic proposals, whereas too large $\varphi$ may introduce a random walk effect or increase momenta rejection rates and thus lessen a potentially positive role of $\varphi$ in tuning sampling performance. 

\begin{figure}[ht!]
	\centering
	\includegraphics[width=0.8\columnwidth ]{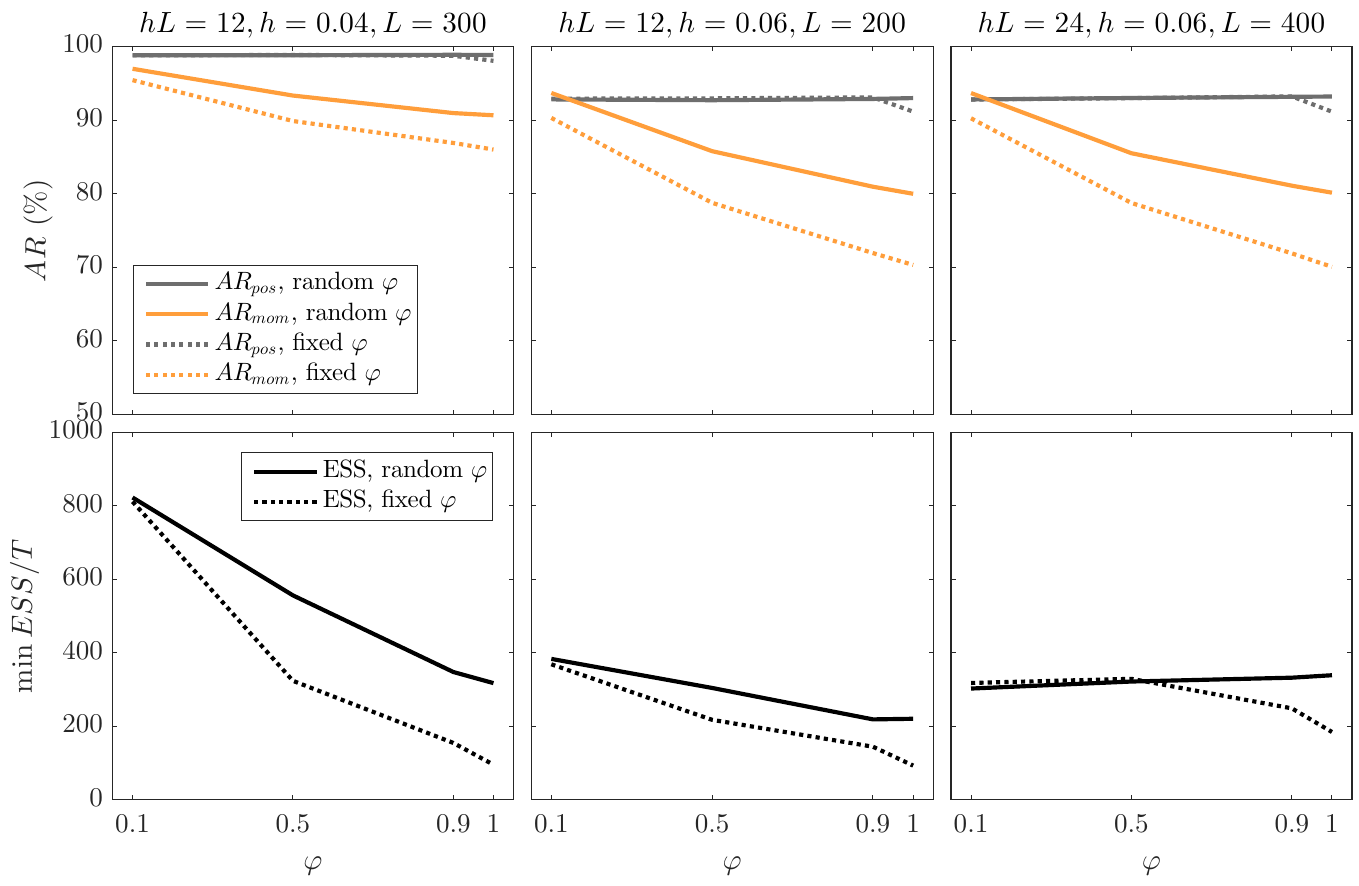} 
	\caption{Position and momenta acceptance rates (top) and time-normalized minimum ESS (bottom)  obtained in sampling from the $100$-dimensional Gaussian distribution using MMHMC with different choices of the trajectory length $hL$ and noise parameter $\varphi$. For each MC iteration, the noise parameter is chosen to be either fixed (dashed line) or random, uniformly drawn from the interval $(0,\varphi)$ (solid line).}
	\label{Fig:phi}
\end{figure}

In Figure \ref{Fig:phi}, we report position and momenta acceptance rates (top) and sampling efficiency, in terms of time-normalized minimum ESS (bottom) in the problem of sampling from the $100$-dimensional Gaussian distribution for different choices of the trajectory length $hL$ and noise parameter $\varphi$. 
Two different schemes for treating the noise parameter $\varphi$ are considered, namely (i) using a fixed value $\varphi$ at every MC iteration, and (ii) choosing a random value uniformly from the interval $(0,\varphi)$.

The figure provides a good illustration of an effect of different parameters of MMHMC on the overall performance of the method. One immediately sees a positive influence of smaller values of the step size $h$ and noise parameter $\varphi$ on the sampling performance of MMHMC. The parameter $L$ seems to play a less important role in the performance tuning. This also applies to the randomization of  $\varphi$ once the optimal value of  $\varphi$ is chosen ($\varphi=0.1$). The situation changes when $\varphi$ is far from its optimal value. In this case the randomization mitigates the effect of those unfavorable choices.  
We summarize the observations specific to a role of $\varphi$ in the MMHMC performance below.

Position acceptance rate is not affected by $\varphi$, unless $\varphi=1$ at which it slightly drops, whereas the acceptance rate of the PMMC step is visibly higher for smaller values of $\varphi$. 
Bigger values of $\varphi$, meaning more random noise introduced in momenta, might stimulate a better space exploration; however, those values lead to more frequent momenta rejections. 
In general, smaller values of $\varphi$ result in better sampling efficiency, though this trend is more obvious for smaller trajectory lengths $hL$.
A noticeable drop in efficiency appears for a fixed value $\varphi=1$, however,  randomization around 1 reduces the negative effect of complete momentum update.

The various numerical tests suggest that a random value from $(0,0.5)$ drawn for every MC iteration is a safe initial guess for a good choice of the parameter $\varphi$. A more theoretically grounded choice of a noise parameter is proposed by \citet{Akhmatskaya:2017}. 

Finally, we note that different values of  $\varphi$ can be assigned to different variates---those that require longer trajectories to decorrelate could have  bigger values of $\varphi$ and those that do not, can use smaller values.

Eventually, an automatic choice of the above free parameters for optimal efficiency can be achieved by adapting the techniques from \citet{WMdF13} to MMHMC \citep{Radivojevic16}.

\paragraph{Order of modified Hamiltonian $k$.} The decision on the order of modified Hamiltonian is not a problematic one. Our experiments indicate that the 4th order modified Hamiltonian combined with the multi-stage integrators performs just well. For more complex models, if the acceptance rate is low with the 4th order, 
the 6th order modified Hamiltonian might be needed. This comes at a higher computational cost; however, such complex models might require large values of $L$ for which the computational overhead due to the calculation of modified Hamiltonian becomes negligible.

For more detailed discussion on the effect of free parameters on the MMHMC performance and accuracy, we refer the reader to \citep{Radivojevic16}. 

\subsubsection{Performance Metrics}\label{Sec:PerfMetrics}

To assess performance of the MMHMC method we use the following metrics:
\begin{itemize}
	\item Acceptance rate (AR); 
	\item Effective Sample Size (ESS) and ESS normalized by the computational time in seconds (ESS/T);
	\item Monte Carlo Standard Error (MCSE) and MCSE normalized by the computational time in seconds (MCSE$\cdot$T);
	\item Efficiency Factor (EF)---relative ESS/T (MCSE$\cdot$T) of MMHMC with respect to	
	another algorithm.
	\item Total distance from the mean, defined as $|| \boldsymbol{\theta}-\boldsymbol{\mu}||=\sum_{d=1}^D|\hat{\theta}_d-\mu_d|$ for the true mean $\boldsymbol{\mu}$, and time-normalized total distance from the mean.
\end{itemize}

\textit{Effective Sample Size}  is a commonly used measure for sampling efficiency of an MCMC method.  It indicates the number of effectively uncorrelated samples out of $N$ collected samples  
and is defined as
\begin{equation*}
ESS_{\text{MCMC}}=\frac{N}{1+2\sum_k \hat\gamma_k},
\end{equation*}
where $\hat\gamma_k$ is the $k$-lag sample autocorrelation \cite{Geyer92}. 

\textit{Monte Carlo Standard Error} of an estimator specifies how much error is in the estimate due to the use of a Monte Carlo method. It is related to  ESS and is defined as
\begin{equation*}
MCSE_{\text{MCMC}}=\sqrt{\frac{\hat{\sigma}^2}{ESS_{\text{MCMC}}}},
\end{equation*} 
where $\hat{\sigma}^2$ is the sample variance.

For general importance sampling methods, high variability in the importance weights might occur if the importance density is not close enough to the target density. In this case ESS is calculated as
\begin{equation}
ESS_{\text{IS}}=\frac{\left(\sum_{n=1}^Nw_n\right)^2}{\sum_{n=1}^Nw_n^2}, \nonumber
\end{equation}
where $w_n,n=1,\dots,N$ are weights associated to all samples, as first introduced by \citet{Kong:1994}.

For importance sampling methods such as GSHMC and MMHMC, one should use a metric for sampling efficiency that takes into account both correlations among samples and weights.
To the best of our knowledge, a metric for samplers that generate correlated weighted samples has not been introduced, though the importance of such an objective criterion was discussed e.g.\ by \citet{Neal98,Gramacy08}.

Here we propose a new metric that addresses these issues and is based on calculation of ESS for  MCMC  and importance samplers jointly. More specifically, we first find the number of uncorrelated samples in the modified ensemble $M:=ESS_{\text{MCMC}}$ using all $N$ posterior samples collected. We estimate ESS$_{\text{MCMC}}$ using the \textsf{CODA} package \cite{Plummer:2006}. Then, we choose $M$ samples out of $N$ by thinning, i.e.\ at a distance of $\lceil N/M\rceil$.
Finally, we calculate MCSE of the importance sampling estimator $\hat I= \sum{}w_nf(\boldsymbol\theta^n)/\sum{}w_n$ for those $M$ uncorrelated samples  as
\begin{equation*}
MCSE_{\text{MCMC-IS}}=\sqrt{\frac{\hat{\sigma}_w^2}{ESS_{\text{MCMC-IS}}}},
\end{equation*}
where  $\hat{\sigma}_w^2$ is the unbiased weighted sample variance \cite{Rimoldini13}
\begin{equation*}
\hat{\sigma}_w^2=\frac{\sum_{n=1}^Mw_n}{(\sum_{n=1}^Mw_n)^2-\sum_{n=1}^Mw_n^2}\sum_{n=1}^Mw_n\left(f(\boldsymbol\theta^n)-\hat{I}\right)^2
\end{equation*}
and
\begin{equation}
ESS_{\text{MCMC-IS}}=\frac{\left(\sum_{n=1}^Mw_n\right)^2}{\sum_{n=1}^Mw_n^2} 
\end{equation}
is the effective sample size for samplers that generate weighted correlated samples.
Note that the effective sample size depends directly on variability in the normalized importance weights. 

Although in the numerical experiments through the paper for MCMC (HMC, GHMC, MALA, RMHMC) and MCMC importance sampling (GSHMC, MMHMC) methods we use the corresponding equations to compute ESS$_{\text{MCMC}}$, ESS$_{\text{MCMC-IS}}$ and MCSE$_{\text{MCMC}}$, MCSE$_{\text{MCMC-IS}}$, we simplify their notation to ESS and MCSE, respectively, in the remainder of the paper.

\subsection{Expected Performance of MMHMC}\label{Sec:PerfMMHMC}

By design, MMHMC incorporates the features and methods, known as potentially favourable for performance enhancement. Among them are irreversibility, importance sampling with modified Hamiltonians (implying high acceptance rates, bounded weights), integration of Hamiltonian dynamics using modified multi-stage splitting integrators (assuring high accuracy and acceptance rates), partial momentum refreshment (resulting in efficient sampling). On the other hand, implementation of such techniques in MMHMC introduces a computational overhead, and using importance sampling may potentially reduce the efficiency of the estimator. Contributions of those factors, positive or negative, into the overall performance of MMHMC are not equivalent, and in this section, we analyze potential performance gains and losses provoked by the most significant factors.   

\begin{figure}[h!]
	\centering
	{\includegraphics[width=0.5\columnwidth ]{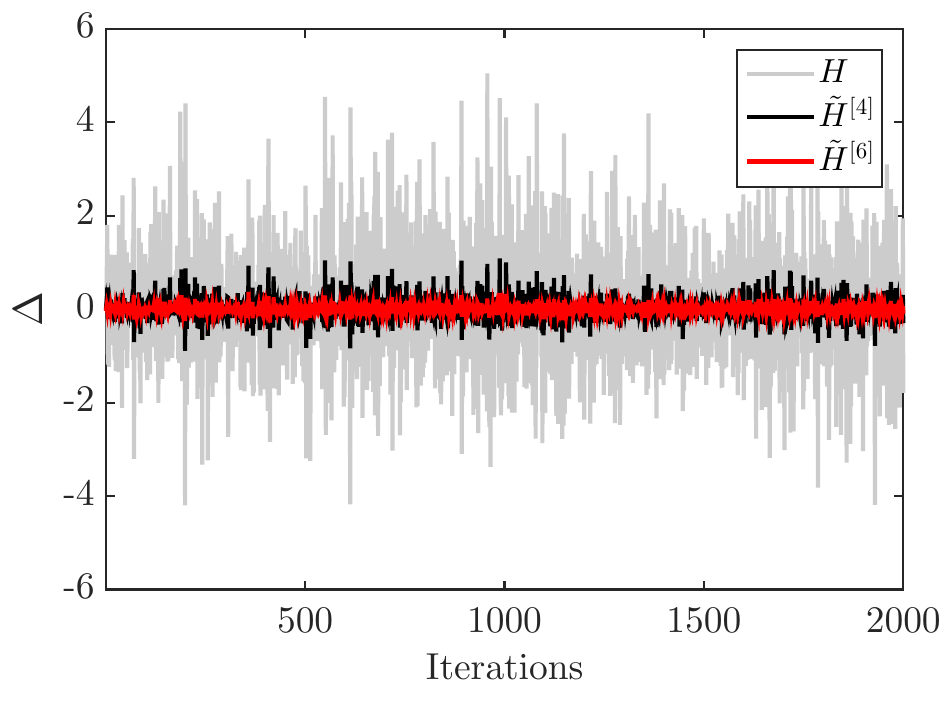} }
	\caption{Observed error in (modified) Hamiltonians after numerical integration with two-stage integrator in MMHMC sampling of a $100$-dimensional Gaussian problem.}
	\label{Fig:Ham_error}
\end{figure}

The main advantage of using an importance distribution defined through modified Hamiltonians comes from the fact that modified Hamiltonians are better preserved by symplectic integrators than true Hamiltonian \cite{LR05}. A better conservation of modified Hamiltonians leads to a smaller error after numerical integration, which directly takes part in the Metropolis test \eqref{HDMetropolis} and results in higher acceptance rates.
For illustration, in Figure \ref{Fig:Ham_error} we compare the resulting numerical integration error $\Delta$ observed in the true Hamiltonian $H$ and the 4th and 6th order modified Hamiltonians given by \eqref{modHam4_an} and \eqref{modHam6_an_Gauss}, respectively, for the $100$-dimensional Gaussian problem. 
$\tilde{H}^{[4]}$ is significantly better conserved than $H$. Conservation of $\tilde{H}^{[6]}$ is even better, as expected. However in practice this must be weighted up against the computational cost of the calculation of the 6th order modified Hamiltonian \eqref{modHam6_an} for general non-Gaussian problems, which includes higher order derivatives. In Section \ref{Sec:Experiments}, we show that the combination of the computationally inexpensive 4th order modified Hamiltonians with accurate multi-stage splitting integrators makes a perfect choice in all numerical experiments, with no need for appealing to higher order expensive modified Hamiltonians.

\begin{figure}[h!]
	\centering
	{\includegraphics[width=0.7\columnwidth]{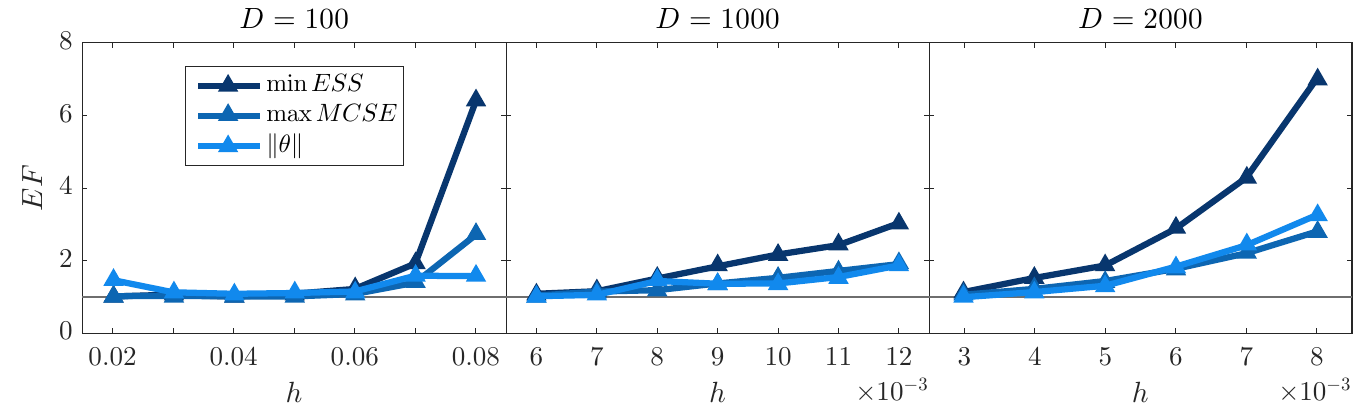} }
	\caption{Relative  efficiency  (EF)  of  MMHMC with fixed step size w.r.t.  MMHMC with randomized step size in  terms  of  minimum ESS, maximum MCSE and total distance from the mean, for a range of step size $h$ and $D$-dimenasional Gaussian model.}
	\label{Fig:rnd_vs_fix}
\end{figure}

Another advantage of using modified Hamiltonians for importance sampling are bounded importance weights ensuring the efficiency of an estimator. Furthermore, avoiding randomization of a step size in MMHMC helps to maintain a low variability of importance weights. Figure \ref{Fig:rnd_vs_fix} demonstrates the superiority of the fixed step strategy over randomization of a step size in MMHMC on the example of the $D$-dimensional Gaussian model.

It may be interesting to compare theoretical performance of HMC and MMHMC for high-dimensional problems. As follows from analysis in Eq.\ \eqref{eq:ExpectationModifiedHamiltonian},
in order to keep acceptance rates in HMC high, an increase in system size $D$ can be counterbalanced by a decrease of a step size $h$ or/and increase in the order $m$ of the symplectic integrator used (\(m \geq 2\), $m=2$ for the Verlet integrator). However, smaller step sizes mean poorer space exploration. This can be partially overcome by increasing a length of the HD trajectory but at the price of reduced computational efficiency. We recall that longer trajectories in HMC imply more frequent time-consuming evaluations of gradients of the potential function. Using high-order symplectic numerical integrators is a possible but rather expensive way of keeping acceptance rates high as such integrators introduce a significant computational overhead. 

For MMHMC (Eq.\ \ref{eq:ExpectationHamiltonian}) 
the order of the modified Hamiltonian \(k \geq 4\)
ensures although shorter than for low-dimensional problems but longer than in HMC, step sizes for high dimensional systems. Moreover, the variability of weights, being a potential threat for MMHMC performance, is lower for smaller time steps, as follows from the definition of the truncated modified Hamiltonian \eqref{eq:TruncatedModHam}.

There are two important reasons for using modified splitting integrators in the MMHMC method. One is their potential to achieve, for a range of step sizes, at a given computational cost, a higher accuracy than Verlet and thus higher acceptance rates and  better space exploration. (We have to emphasize that for fair comparison, different integrators have to be applied with the same computational effort, rather than with the same step length; an $r$-stage integrator requires $r$ gradient evaluations per time step and to be compared with Verlet has to be used with a step length correspondingly longer.)

A second possible benefit of the integrators of this class is that, due to the extra accuracy, they may avoid the need for computationally expensive, higher order modified Hamiltonians.

Numerical experiments confirm that the Verlet integrator currently used within HMC and MHMC methods can be advantageously replaced in MMHMC with modified multi-stage integrators whose implementation is essentially that of Verlet \citep{Radivojevic:2018}. The modified two- and three-stage integrators lead to an outstanding improvement (up to 8 times) over Verlet in terms of acceptance rate and sampling efficiency, for a range of step sizes, for high dimensional problems in which the potential function is (approximately) quadratic. 

An introduction of the modified partial momentum update in MMHMC intends to reduce a computational overhead caused by the evaluation of modified Hamiltonians within the Metropolis test in the PMMC step. 
The proposed PMMC step is at least as efficient as the original momentum update implemented in GSHMC, whereas for specific choices of models and parameters it may demonstrate a far better computational performance that can be achieved with the original algorithm.

\begin{figure}[t!]
	\centering
	{\includegraphics[width=0.5\columnwidth ]{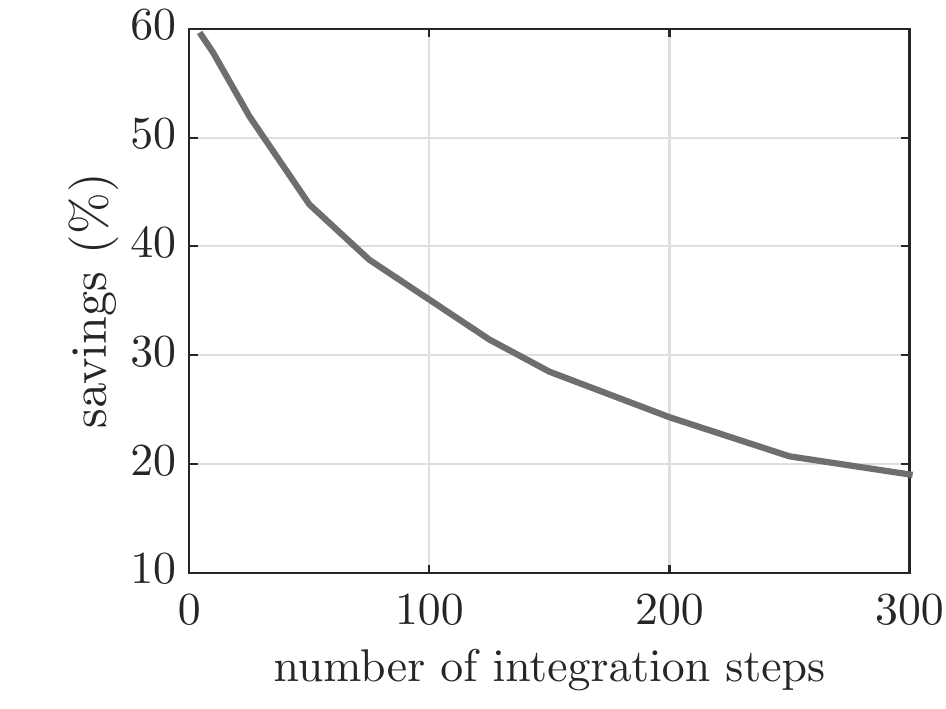} }
	\caption{Savings in computational time observed in MMHMC sampling of a model with
		dense Hessian matrix after replacing the original PMMC step with the newly proposed PMMC step. The 4th order modified Hamiltonian \eqref{modHam4_an} with analytical derivatives was used.}
	\label{Fig:newPMMC}
\end{figure}

In Figure \ref{Fig:newPMMC} we show the savings in computational time observed in MMHMC sampling of a model with a dense Hessian matrix after replacing the original PMMC step with the newly proposed one. The modified Hamiltonian \eqref{modHam4_an} and the range of HD trajectories lengths have been considered in this case.  Clearly, the new PMMC step improves the efficiency of MMHMC in sampling such models (up to 60\%), especially if moderately short HD trajectories, favoured in MMHMC, are chosen.  

\begin{figure}[h!]
	\centering
	{\includegraphics[width=0.45\columnwidth ]{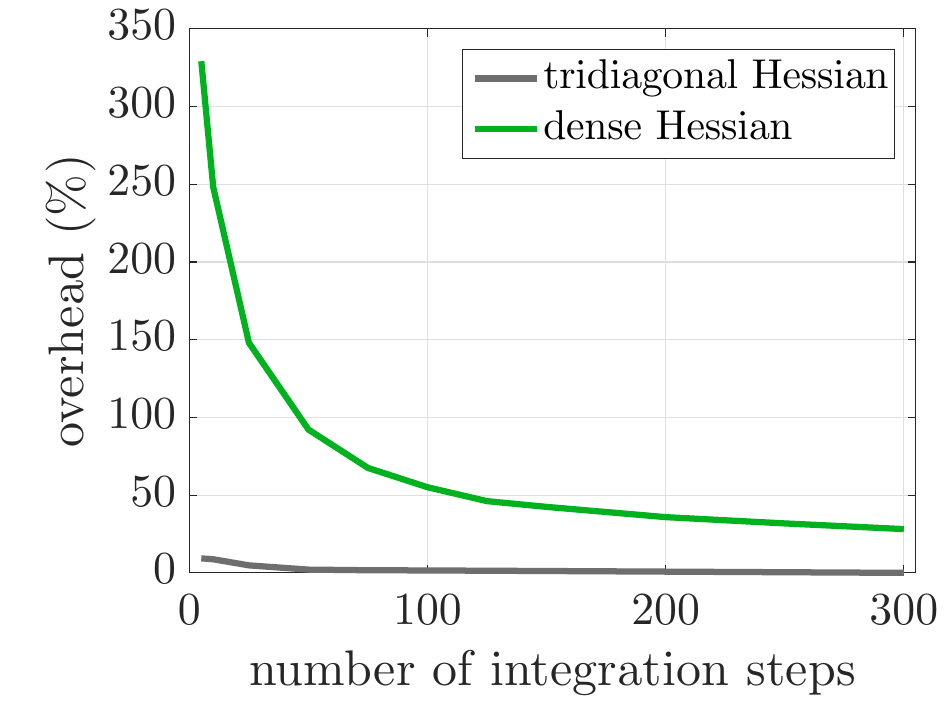} }
	{\includegraphics[width=0.45\columnwidth ]{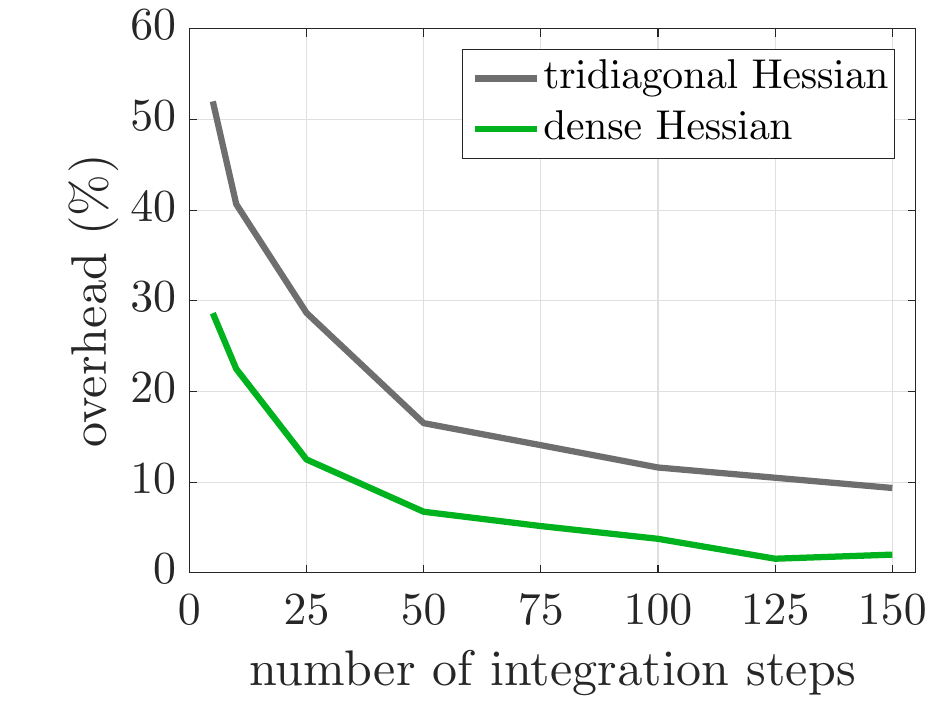} }
	\caption{Computational overhead of MMHMC compared to HMC for models with a tridiagonal and a dense Hessian matrix using the 4th order modified Hamiltonian \eqref{modHam4_an} with all required derivatives calculated analytically (left), and the 4th order modified Hamiltonian \eqref{modHam4num} with numerical approximation of the time derivatives (right).}
	\label{Fig:overheads}
\end{figure}

The computational effort required for calculation of modified Hamiltonians in MMHMC is the crucial issue for the overall performance efficiency of MMHMC. In general, the higher orders modified Hamiltonians are more computationally demanding than the ones of the low orders. For models with a tridiagonal Hessian matrix, the modified Hamiltonians with analytical derivatives \eqref{modHam4_an}--\eqref{modHam6_an_Gauss} introduce less computational overhead than those expressed in terms of numerical time derivatives \eqref{modHam4num}-\eqref{modHam6num}, whereas for models with a dense Hessian matrix, the modified Hamiltonians \eqref{modHam4num}-\eqref{modHam6num} are less expensive than \eqref{modHam4_an}--\eqref{modHam6_an_Gauss}. As stated before, avoiding modified Hamiltonians of orders higher than 4 became possible with the introduction in MMHMC of accurate modified splitting integrators specifically tuned for the MHMC methods.
Figure \ref{Fig:overheads} shows computational overheads of MMHMC, compared to the HMC method, for models with tridiagonal and dense Hessian matrices when MMHMC uses the 4th order modified Hamiltonian with derivatives calculated analytically \eqref{modHam4_an} (left panel), and the 4th order modified Hamiltonian with numerical approximation of the time derivatives \eqref{modHam4num} (right panel). 
Figure \ref{Fig:overheads} (left)  illustrates that models with dense Hessian matrices imply non-negligible overhead. In all other cases, the overheads are minor unless the number of integration steps becomes very small.

Dependence of MMHMC performance on a choice of tunable parameters is yet another factor which may deteriorate MMHMC efficiency. This is a well-known drawback common to all HMC-based methods. The advantage of vanilla HMC over other HMC methods discussed in this section comes from a fewer number of parameters to tune, due to an absence of partial momentum update in its algorithm.     

In the final analysis, in Table \ref{MMHMC_features} we summarize the differences between four somewhat similar methods, MMHMC, HMC, GHMC, GSHMC, in terms of how the presence or absence of various MMHMC features affects their capacity to sample efficiently.

We excluded randomization of methods' parameters from Table \ref{MMHMC_features} since its impact on performance is inconsistent. While randomization of parameters normally improves performance in HMC, a randomized step size in MMHMC leads to an opposite effect, as was discussed above. Moreover, the GSHMC method has been formulated for physical applications where parameters have physical meaning and are assumed to be fixed by default.

\setlength{\tabcolsep}{1pt}
\begin{table}[ht]
	\caption{Presence of performance impacting factors in HMC-based algorithms. (Bold symbols imply higher impacts) }
	\begin{center}
		\begin{tabular}{c c c c c}
			\toprule
			\multicolumn{5}{c}{Performance Enhancement}\\
			\midrule
			& MMHMC& HMC & GHMC & GSHMC\\ 
			Irreversibility & yes & no & yes & yes \\
			Modified Hamiltonians & yes& no& no& yes  \\ 
			PMMC & yes& no& yes& yes \\
			Splitting Integrators & yes& no& no& no \\
			\midrule
			\multicolumn{5}{c}{Performance Degradation}\\
			\midrule
			{Computation of High Order}  & \multirow{ 2}{*}{yes}& \multirow{ 2}{*}{no}& \multirow{ 2}{*}{no}& \multirow{ 2}{*}{\textbf{yes}}   \\ 
			Derivatives & & & & \\
			Variability of Weights & yes& no& no & yes \\ 
			Ambiguous Choice  
			& \multirow{ 2}{*}{\textbf{yes}}& \multirow{ 2}{*}{yes}& \multirow{ 2}{*}{\textbf{yes}} & \multirow{ 2}{*}{\textbf{yes}} \\
			of Parameters  & & & & \\
			\bottomrule
		\end{tabular}
		\label{MMHMC_features}
	\end{center}
\end{table}

\section{Numerical Experiments}\label{Sec:Experiments}

In this section we examine the performance of MMHMC on various benchmark models and compare it against other popular sampling techniques in computational statistics to answer the question of whether MMHMC emerges as a competitor to the most successful methods like HMC and RMHMC.

\subsection{Implementation}\label{Sec:Implementation}

The MMHMC method has been implemented in the user-friendly in-house software package \textsf{HaiCS} (Hamiltonians in Computational Statistics), written in \textsf{C} and targeted to computers running UNIX certified operating systems. 

The code is intended for statistical sampling of high dimensional and complex distributions and parameter estimation in different models through Bayesian inference using Hamiltonian Monte Carlo based methods. The currently available sampling techniques include the Metropolis algorithm, Hamiltonian Monte Carlo (HMC), Generalized Hybrid Monte Carlo (GHMC),
Metropolis Adjusted Langevin Algorithm (MALA), second order Langevin Monte Carlo (L2MC), Generalized Shadow Hybrid Monte Carlo (GSHMC) and Mix \& Match Hamiltonian Monte Carlo (MMHMC), the method presented in this paper. 
The package benefits from efficient implementation of modified Hamiltonians, the accurate multi-stage splitting integration schemes
, the analysis tools compatible with \textsf{CODA} toolkit for MCMC diagnostics as well as an interface for implementing alternative splitting integrators and complex statistical models. The popular statistical models, such as, multivariate Gaussian distribution, Bayesian Logistic Regression and Stochastic Volatility are implemented in \textsf{HaiCS}.

The complete description of \textsf{HaiCS} package can be found in \citep{Radivojevic16}. 

\subsection{Experimental Results}

We evaluate the performance of the MMHMC method and compare it with the Random Walk Metropolis-Hastings (RWMH), Hamiltonian Monte Carlo (HMC), Generalized Hybrid Monte Carlo (GHMC), Metropolis Adjusted Langevin Algorithm (MALA), Riemann Manifold HMC (RMHMC) and Generalized Shadow Hybrid Monte Carlo (GSHMC) methods on a set of standard benchmark models used in the literature. Space exploration and/or sampling efficiency are examined on the banana-shaped distribution, multivariate Gaussian distribution, Bayesian logistic regression model, and the stochastic volatility model.

The choice of the optimal parameters of the algorithms remains an open question \cite{Neal10} and not the subject of this paper. To make the comparison with other methods fair, we chose the following strategy. Since the stochastic volatility  benchmark is studied well in literature, and HMC and RMHMC were tuned previously for a particular dimension of this benchmark, we took the found sets of optimal parameters as an initial guess and tuned them further. For Bayesian logistic regression and Gaussian model, especially for some data sets, such information is not available. In this case, we have located a range of reasonable parameters $L,h$ and $\varphi$ and performed the comparison for these sets. 

For each MC iteration we draw the number of integration steps uniformly from $\{1,\dots,L \}$ for HMC and GHMC, and step size uniformly from $(0.8h, 1.2h)$ for HMC, GHMC and MALA methods. For GSHMC, we hold all parameters fixed as originally proposed in the method. 
Naturally, for $r$-stage integrators, a step size is set to $rh$ and a number of integration steps to $L/r$. 
We observed that bigger values of $L$ yield higher efficiency for HMC and GHMC for all tested step sizes, whereas for GSHMC and MMHMC this is not the case.
Additionally, we tested MMHMC for a range of noise parameters $\varphi$ being fixed as well as drawn uniformly from $(0,\varphi)$. Smaller values of $\varphi$  tend to perform better for smaller values of the product $hL$ and vice versa. Nevertheless, here we report only results obtained with the best $\varphi$ and $L$ among tested for each step size $h$. 
Complete experimental setup for each method and model tested is given in Appendix \ref{Sec:Exp_setup}. 
All our experiments are carried out with {\color{black}the identity mass matrix} for HMC, GHMC, MALA, GSHMC and MMHMC.  

In the results presented here, we compute ESS (MCSE) of the mean estimator for each variate, as proposed in \ref{Sec:PerfMetrics}, and report minimum, median, and maximum ESS (MCSE) across variates or just minimum ESS (maximum MCSE), as the most restrictive measures, calculated using the collected posterior samples.
Computational time used for normalization of ESS, MCSE and efficiency comparison is measured as CPU time that each method takes to collect posterior samples.
Except for the case of the banana-shaped distribution, for which we investigate a typical trajectory of a single Markov chain, all results are averaged over ten independent runs. 

We examine the banana-shaped model with the Matlab code provided along with the paper by \citet{Lan:2012}, in which we implemented the MMHMC method. The rest of experiments are carried out with the in-house software package \textsf{HaiCS}, outlined in Section \ref{Sec:Implementation}.

Each test model has been prepared to sampling with MMHMC, which in the first instance involved computation of derivatives of a model potential function. 

\subsubsection{Banana-shaped Distribution}
We begin with a comparison of a space exploration achieved by MMHMC,  RWMH, HMC and RMHMC in sampling of a  2-dimensional, non-linear target. 
The idea is to illustrate a representative mechanism of exploring a space for each tested method by generating a typical trajectory of a single Markov chain.
Given data $\mathbf y=\{y_k\}_{k=1}^K$ we sample from the banana-shaped  posterior distribution of the parameter  $\boldsymbol{\theta}=(\theta_1,\theta_2)$  \citep{BC11} for which the likelihood and prior distributions are given as
\begin{eqnarray*}
	y_k|\boldsymbol{\theta}&\sim&\mathcal{N}(\theta_1+\theta_2^2,\sigma^2_{y}), \; k=1,\dots,K,\\
	{\theta}_1,\theta_2  &\sim& \mathcal{N}(0,\sigma^2_{\theta}),
\end{eqnarray*}
respectively.
Due to independency in the data and parameters, the posterior distribution $\pi(\boldsymbol{\theta}|\mathbf y)$ is proportional to
$$\prod_{k=1}^K p(y_k|\boldsymbol{\theta}) p(\theta_1)p(\theta_2).$$

\paragraph{Experimental setting.} Data $\{y_k\}_{k=1}^K$, $K=100$ are generated with $\theta_1+\theta_2^2=1$,  $\sigma_{y}=2$ and $\sigma_{\theta}=1$. 
Sampling with the MMHMC method is performed using the Verlet integrator and the modified Hamiltonian \eqref{modHam4_an}, a fixed number of integration steps, a step size and a noise parameter with values $L=7, h=1/9, \varphi=0.5$, respectively. MMHMC is compared with RWMH, HMC and  RMHMC for which simulation parameters are chosen as suggested by \citet{Lan:2012}. 

\begin{figure}[ht!]
	\centering
	{\includegraphics[width=0.75\columnwidth ]{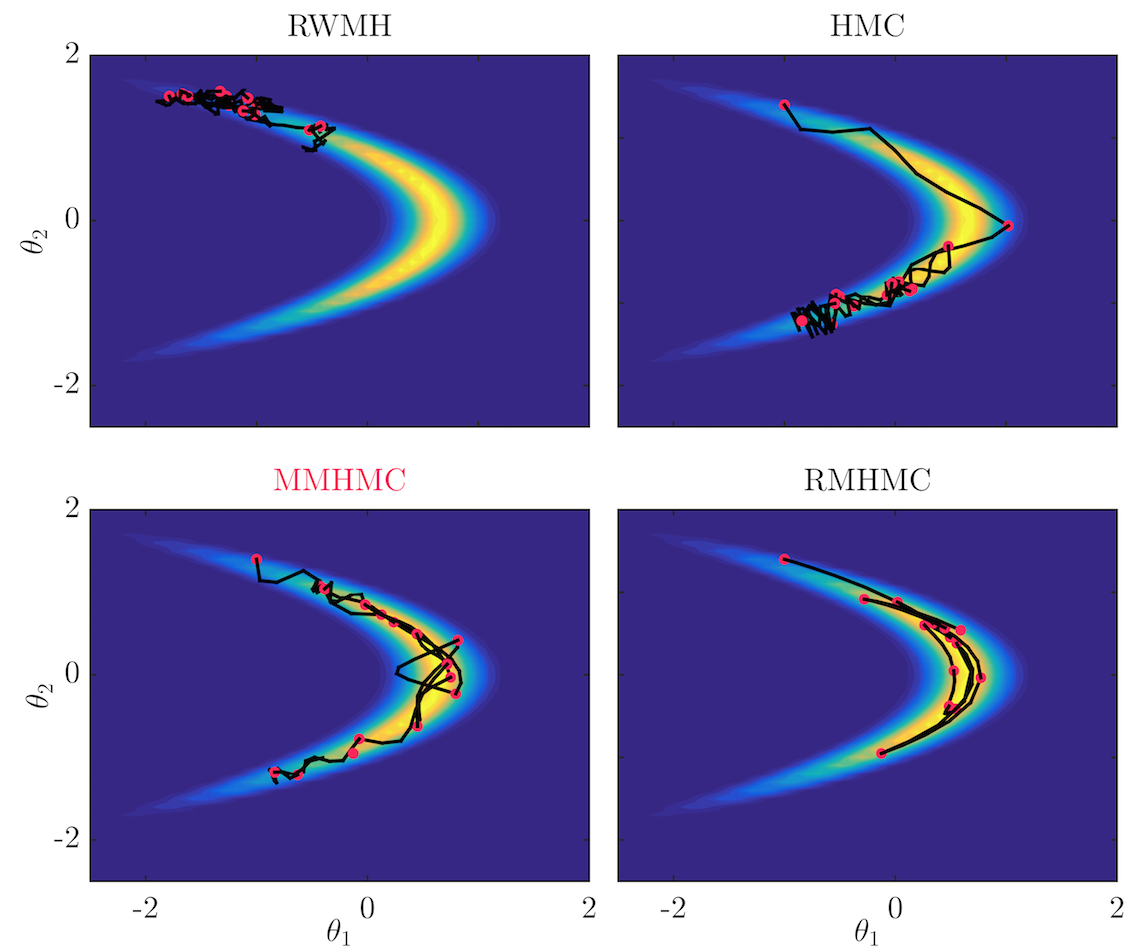} }
	\caption{The first 15 Monte Carlo iterations with sampling paths (lines) and accepted proposals (dots) in sampling from the banana-shaped distribution with Random Walk Metropolis-Hastings (RWMH), Hamiltonian Monte Carlo (HMC), Mix \& Match HMC (MMHMC) and Riemann Manifold HMC (RMHMC).}
	\label{Fig:Banana_paths}
\end{figure}

\paragraph{Results.} The dynamics of the four samplers is illustrated in Figure \ref{Fig:Banana_paths}, in which sampling paths (lines) of the first 15 accepted proposals (dots) are shown. 
RWMH just has started to explore the parameter space and is still located in the low-density tail. In contrast, other methods already have visited high-density regions.
As expected, RMHMC efficiently tracks a local curvature of the parameter space and is able to move along the ridge to its full extent.
On the other hand, HMC and MMHMC  tend to move across rather than along the ridge, with MMHMC sampling visibly broader than does HMC.

\subsubsection{Multivariate Gaussian Distribution}
This benchmark has been proposed by \citet{HoffmanGelman14}. The task is to  sample from the $D$-dimensional Gaussian $\mathcal N(0,\Sigma)$, where the precision matrix $\Sigma^{-1}$ is generated from a Wishart distribution with $D$ degrees of freedom and the $D$-dimensional identity scale matrix.

\paragraph{Experimental setting.} The tests are performed for three different dimensions, $D=100,1000,2000$, using the HMC, GHMC, GSHMC and MMHMC methods. 
For the identity mass matrix, all four methods are invariant under rotations. Therefore, due to limited computational resources, for cases $D=1000, 2000$ we choose the covariance matrix $\Sigma$ to be diagonal with 
$$\Sigma_{ii}=\sigma^2_i,$$
where $\sigma^2_i$ is the $i$th smallest eigenvalue of the original covariance matrix. 
Sampling with MMHMC is performed using the modified Hamiltonian \eqref{modHam4_an}, and the M-BCSS3 and M-ME3 integrators for $D=100$ and $D=1000,2000$, respectively. {\color{black}For simplicity, we use the same formulation and implementation of the modified Hamiltonian in GSHMC as in MMHMC. However, we notice that in the original GSHMC algorithm the less efficient implementation of the modified Hamiltonian is proposed and thus the GSHMC performance in the following tests is likely overestimated.}   
$10000,20000,30000$ samples are collected with each method with first $2000$, $5000$, $5000$ being discarded as a warm-up for dimensions $D=100,1000,2000$, respectively. 

\begin{figure}[ht!]
	\centering
	\includegraphics[width=0.85\columnwidth ]{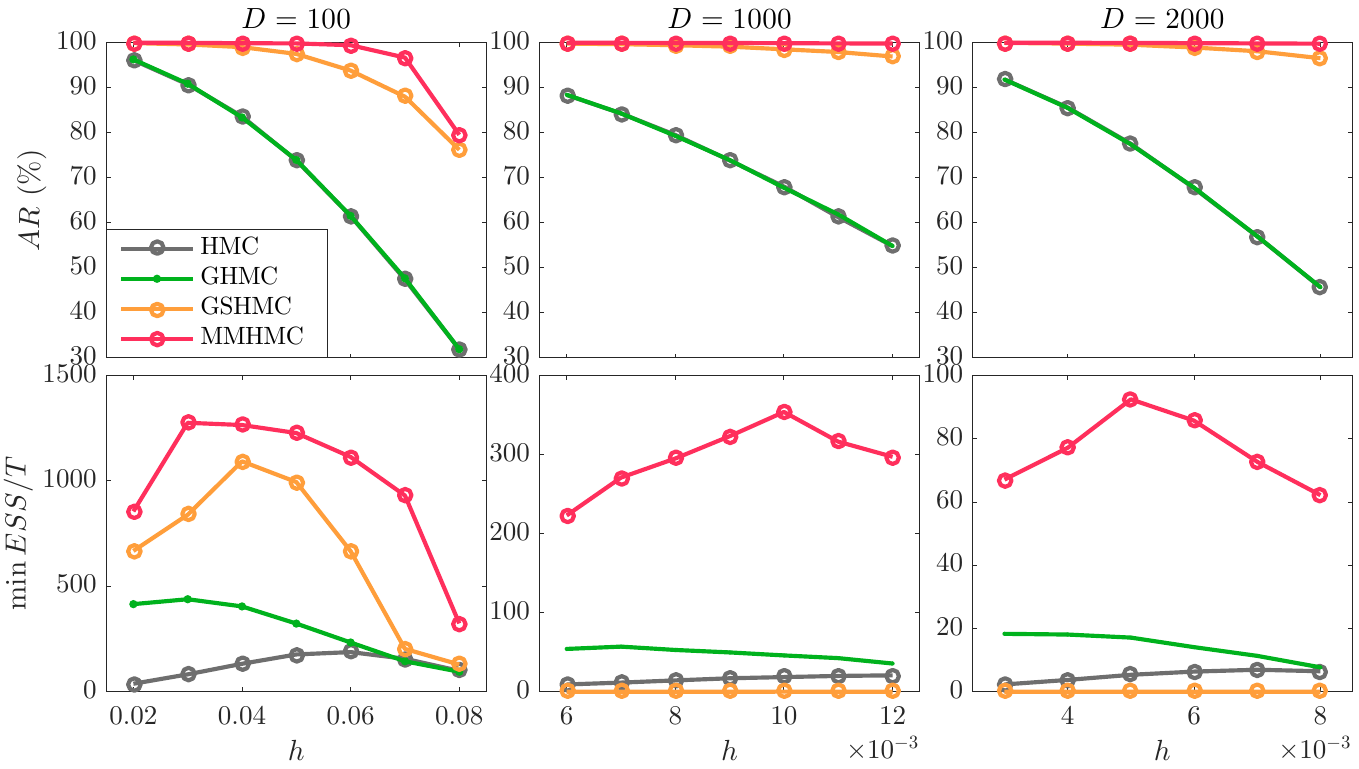} 
	\caption{$D$-dimensional Gaussian distribution. Acceptance rate (top) and time-normalized minimum ESS (bottom) for a range of step sizes $h$, obtained in sampling with Hamiltonian Monte Carlo (HMC), Generalized Hybrid Monte Carlo (GHMC), Generalized Shadow Hybrid Monte Carlo (GSHMC) and Mix \& Match HMC (MMHMC).}
	\label{Fig:Gauss_AR+ESS} 
\end{figure}

\begin{figure}[ht!]
	\centering
	\includegraphics[width=0.85\columnwidth ]{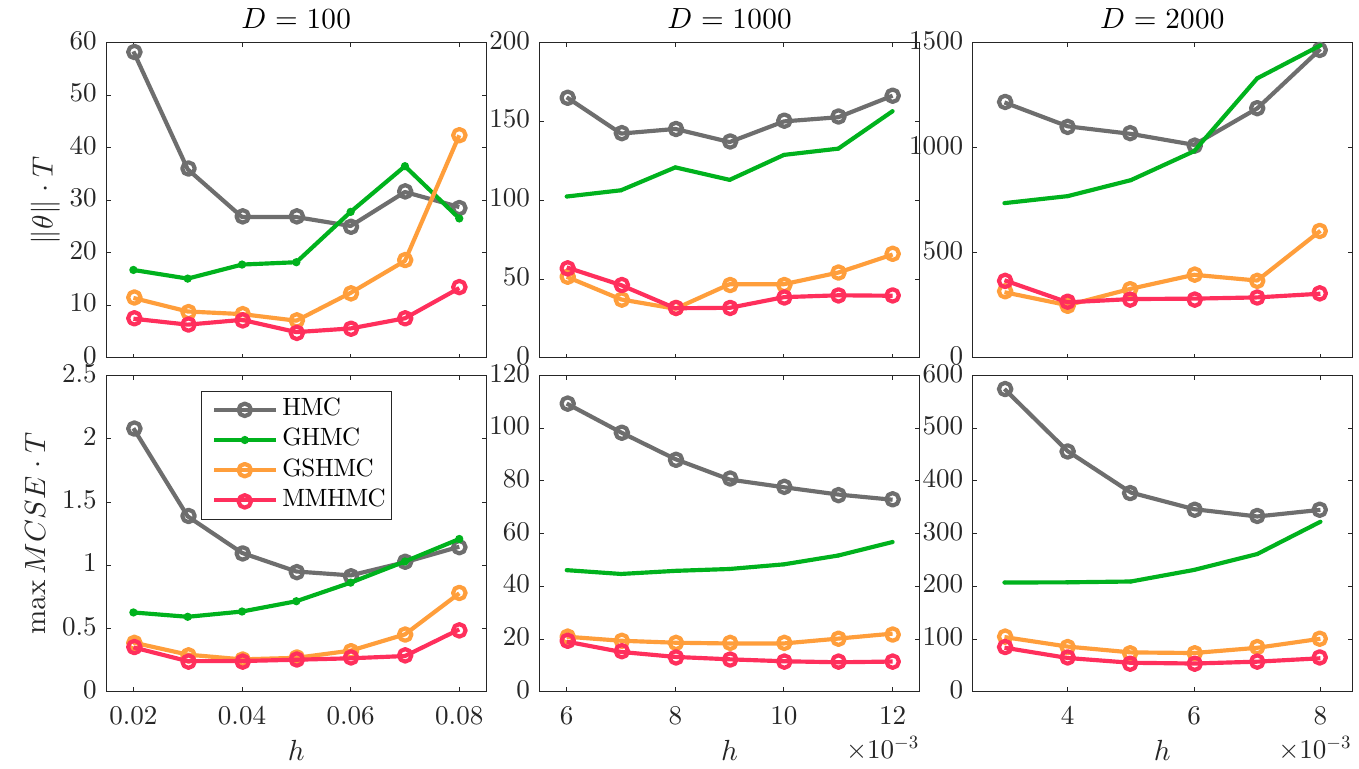} 
	\caption{$D$-dimensional Gaussian distribution. Time-normalized total distance from the mean (top) and maximal MCSE (bottom) for a range of step sizes $h$, obtained in sampling with Hamiltonian Monte Carlo (HMC), Generalized Hybrid Monte Carlo (GHMC), Generalized Shadow Hybrid Monte Carlo (GSHMC) and Mix \& Match HMC (MMHMC).}
	\label{Fig:Gauss_Dist+MCSE} 
\end{figure}

\paragraph{Results.}
Figure \ref{Fig:Gauss_AR+ESS} compares the obtained acceptance rates (top) and corresponding time-normalized minimum ESS (bottom).
While acceptance rates for HMC and GHMC drop considerably with increasing step size, especially for higher dimensions, MMHMC, in particular, and GSHMC maintain very high acceptance.
For $D=100$ acceptance rates for MMHMC and GSHMC start to drop visibly but still stay reasonably high for longest step sizes. 
In addition, Figure \ref{Fig:Gauss_Dist+MCSE} presents the comparison in terms of time-normalized total distance from the mean $\|\boldsymbol{\theta}\|$ (top), and maximal MCSE (bottom) obtained with the four methods, where lower values correspond to better performance.
As can be seen from the inspection of time-normalized ESS, MCSE and $\|\boldsymbol{\theta}\|$, for all tests, MMHMC 
outperforms in sampling efficiency all considered methods.


\begin{figure}[ht!]
	\centering
	\includegraphics[width=0.85\columnwidth ]{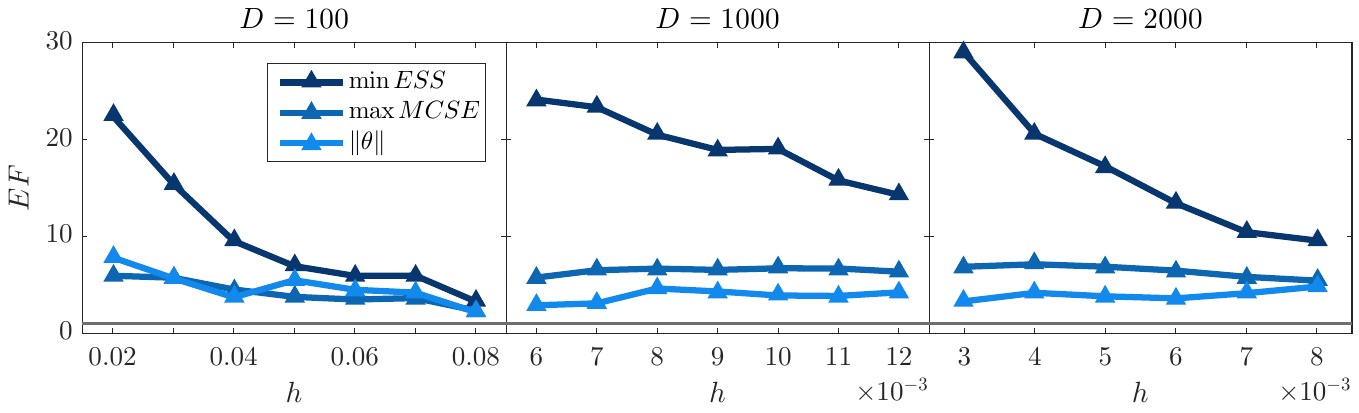} 
	\caption{$D$-dimensional Gaussian distribution. Relative efficiency (EF) of MMHMC w.r.t.\ HMC in terms of time-normalized minimum ESS, maximum MCSE and total distance from the mean, for a range of step sizes $h$.}
	\label{Fig:Gauss_Ef} 
\end{figure}

The results on sampling efficiency are summarized in Figure \ref{Fig:Gauss_Ef}, from which one can appreciate the amount of improvement achieved with MMHMC compared to HMC. For a range of step sizes $h$ the efficiency factor (EF) in terms of time-normalized minimum ESS, maximum MCSE and total distance, relative with respect to HMC, is shown in such a way that  values above 1 indicate superior performance of MMHMC. 
The improvement factor slowly increases with dimension. Depending on the choice of $h$, starting from at least a comparable performance (for the lowest dimension), the maximal improvement goes up to 29 times (for the highest dimension). 

\begin{figure}[ht!]
	\centering
	\includegraphics[width=0.85\columnwidth ]{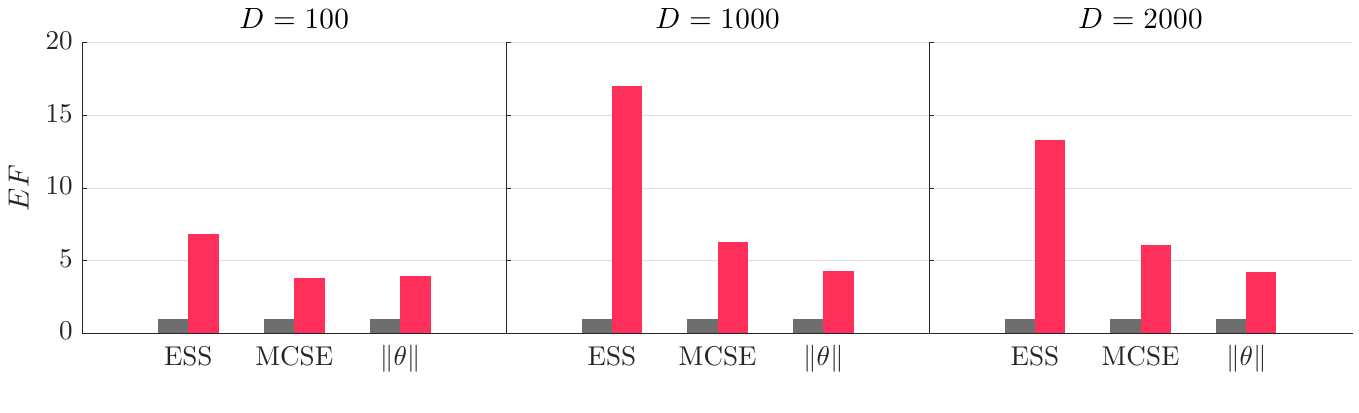} 
	\caption{$D$-dimensional Gaussian distribution. Relative efficiency (EF) of MMHMC w.r.t.\ HMC in terms of time-normalized minimum ESS, maximum MCSE and total distance from the mean, achieved using the best set of parameters for each method.}
	\label{Fig:Gauss_best} 
\end{figure}

Finally, Figure \ref{Fig:Gauss_best} summarizes the improvements obtained with MMHMC compared to HMC in terms of the same metrics, when considering the results achieved with the best set of parameters for each method and each dimension found among the tested ones. Clearly, the MMHMC method demonstrates superiority for all the three metrics considered, especially in terms of ESS. However, in a general case, the optimal parameters are not known a priori for either of the sampling methodologies.

\subsubsection{Bayesian Logistic Regression Model}
The Bayesian logistic regression (BLR) model  is used for solving binary classification problems appearing across various fields such as medical and social sciences, engineering, insurance, ecology, sports, etc.

Let consider $K$ instances of data $\{\mathbf x_k, y_k\}_{k=1}^K$, where $\mathbf x_k$ are vectors of $D-1$ covariates and $y_k \in \{0,1\}$ are binary responses. 
In the BLR model, response variable $\mathbf y=(y_1,\dots,y_K)$ is governed by a Bernoulli distribution with a parameter $\mathbf p=(p_1,\dots,p_K)$.
The unobserved probability  $p_k$ of a particular outcome is linked to the linear predictor function through the logit function, i.e. 
$$\textrm{logit}(p_k)=\theta_0 +\theta_1x_{1,k}+\cdots+\theta_{D-1}x_{D-1,k},$$
where  $\boldsymbol{\theta} \in \mathbb R^D$ is the regression coefficient vector. The prior of the regression coefficient can be chosen e.g.\ as $\boldsymbol{\theta}\sim \mathcal{N}(0,\alpha \mathbb I)$, with a known $\alpha$. 

If we construct the design matrix $X \in \mathbb R^{K,D}$ of input data as
$$X=\begin{bmatrix}
1 & x_{11} & \cdots& x_{1,D-1} \\
\vdots & \vdots & & \vdots \\
1 & x_{K1} & \cdots& x_{K,D-1}
\end{bmatrix},$$
the likelihood function is given as
\begin{equation*}
\begin{aligned}
p(\mathbf y|X,\boldsymbol{\theta})&=\prod_{k=1}^K p(y_k|X_k,\boldsymbol{\theta})\\
&=\prod_{k=1}^K \left( \frac{ \mathrm{e}^{X_k\boldsymbol{\theta}} }{ 1+\mathrm{e}^{X_k\boldsymbol{\theta}} } \right)^{y_k} \left(\frac{1}{1+\mathrm{e}^{X_k\boldsymbol{\theta}}} \right)^{1-y_k},
\end{aligned}
\end{equation*}
where $X_k$ is the $k$th row of the matrix $X$.
The corresponding posterior distribution over the regression coefficients is
$$\pi(\boldsymbol{\theta}|\mathbf y,\mathbf x)\propto \prod_{k=1}^K p(y_k|X_k,\boldsymbol{\theta}) p(\boldsymbol\theta)$$
with the prior
$$p(\boldsymbol{\theta})\propto\exp\left\{ -\frac{\boldsymbol{\theta}^T\boldsymbol{\theta}}{2\alpha}\right\}.$$

\paragraph{Experimental setting.} 
We use four different real data sets available from the University of California Irvine Machine Learning Repository \cite{Lichman13}. The data set characteristics, such as names, numbers of regression parameters ($D$) and observations ($K$) are summarized in Table \ref{BLR_datasets}. 
\begin{table}
	\caption{Data sets used for the BLR model with corresponding numbers of regression parameters ($D$) and numbers of observations ($K$).}
	\centering
	\begin{tabular}{ccc}
		\toprule
		Data set&$D$& $K$\\ 
		\midrule
		German&25 &1000\\
		Sonar&61&208\\
		Musk&167&476\\  
		Secom&444&1567\\
		\bottomrule
	\end{tabular}
	\label{BLR_datasets}
\end{table}

By following a common procedure, we normalize input data such that each covariate has zero mean and standard deviation of one.
For each data set, a diffuse Gaussian prior is imposed by setting $\alpha=100$. 

For the German and Sonar data sets, $N = 5000$ posterior samples were generated after discarding the first $1000$ samples as a warm-up, while for the bigger data sets (Musk and Secom) twice as much samples were collected.
Apart from the comparison of MMHMC with HMC over the range of data sets, we also tested it against MALA on German data set.
We do not investigate the performance of RMHMC since, as it was stated by \citet{GC11}, RMHMC does not outperform HMC for dimensions as high as for the German data set ($D=25$), which in our case is the data set of the smallest dimension. 

In these experiments, MMHMC is used with the modified Hamiltonian \eqref{modHam4num} and the Verlet integrator.

\begin{figure}[h!]
	\centering
	\includegraphics[width=0.65\columnwidth ]{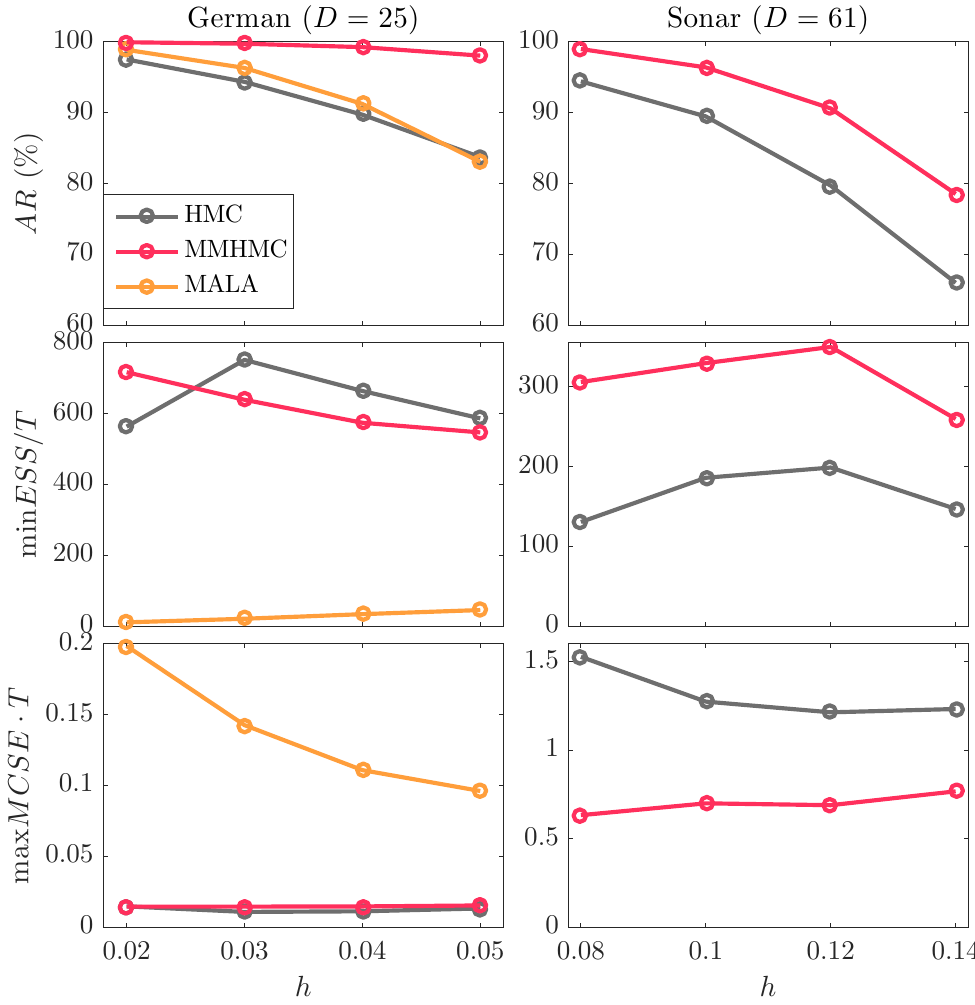} 
	\caption{Bayesian logistic regression. Acceptance rate (top), time-normalized minimum ESS (middle) and maximum MCSE (bottom) across variates obtained using Hamiltonian Monte Carlo (HMC), Mix \& Match HMC (MMHMC) and Metropolis Adjusted Langevin Algorithm (MALA), for a range of step sizes $h$, for the German and Sonar data sets.}
	\label{BLR_AR_ESS_gs}
\end{figure}

\begin{figure}[h!]
	\centering
	\includegraphics[width=0.65\columnwidth ]{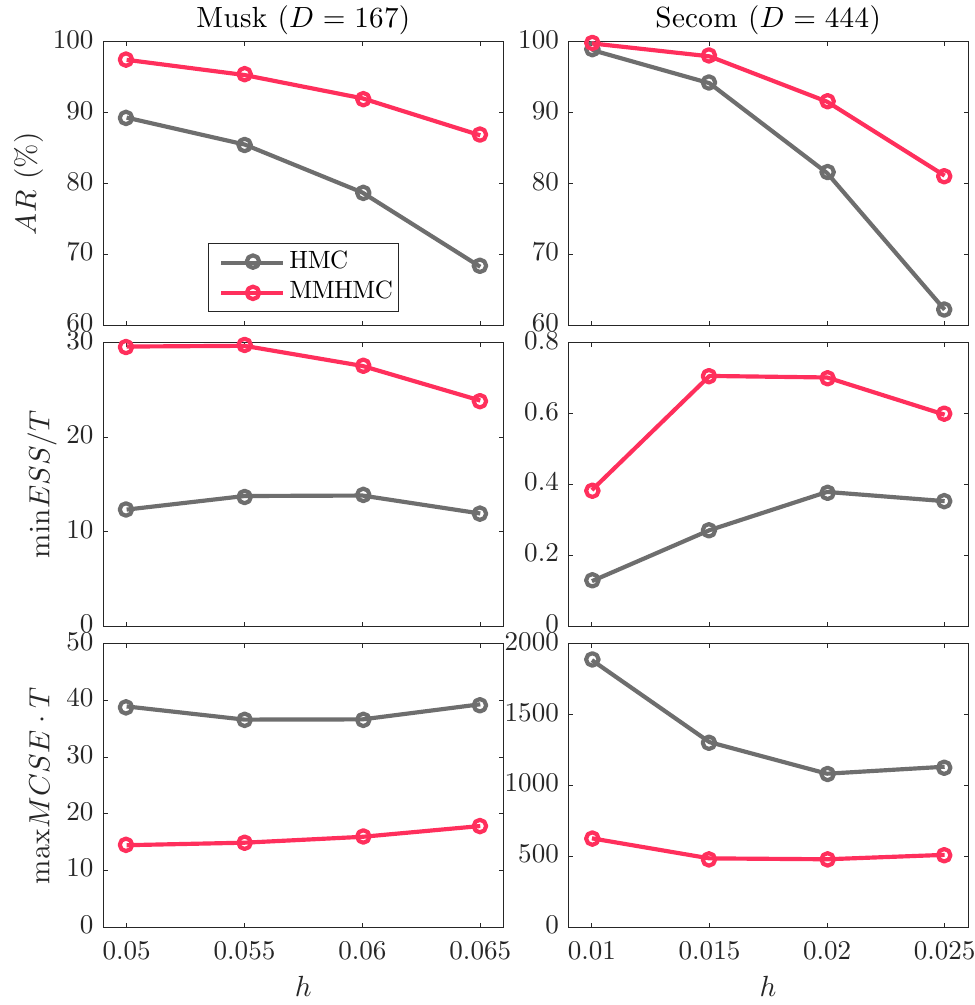} 
	\caption{Bayesian logistic regression. Acceptance rate (top), time-normalized minimum ESS (middle) and maximum MCSE (bottom) across variates obtained using HMC and MMHMC, for a range of step sizes $h$, for the Musk and Secom data sets.}
	\label{BLR_AR_ESS_ms}
\end{figure}

\paragraph{Results.}
Acceptance rate (top), time-normalized minimum ESS (middle) and maximum MCSE (bottom) across variates obtained for BLR are presented in Figures \ref{BLR_AR_ESS_gs} and \ref{BLR_AR_ESS_ms}. 
For all data sets, acceptance rate is the highest for MMHMC, as expected. 
For the smallest data set, while MALA exhibits visibly poor performance, both HMC and MMHMC demonstrate high and comparable efficiency. 
The trend changes for HMC method with increasing size of a problem. The superiority of MMHMC over HMC becomes more noticeable when a bigger data set is considered, resulting in the performance improvement by a factor of over 3 for the Secom data set ($D = 444$).  

\begin{figure}[h!]
	\includegraphics[width=1\columnwidth ]{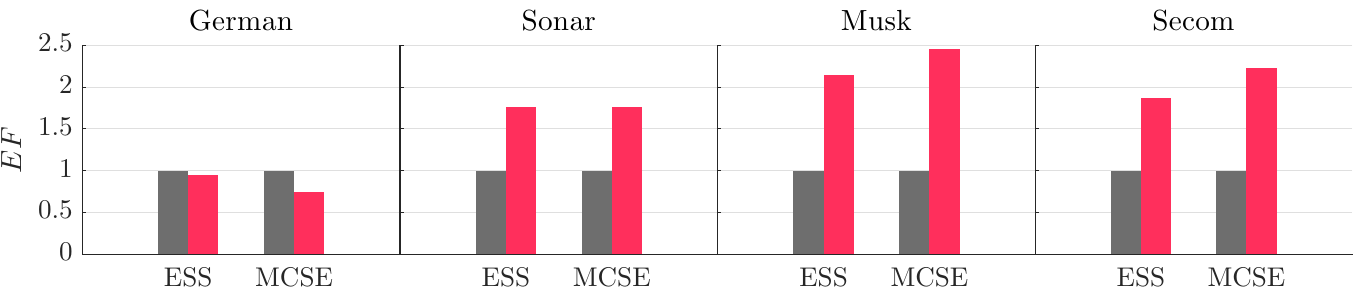} 
	\caption{Bayesian logistic regression. Relative efficiency (EF) of MMHMC w.r.t.\ HMC in terms of time-normalized minimum ESS and maximum MCSE across variates achieved using the best set of simulation parameters for each method.}
	\label{BLR_EF}
\end{figure}

Figure \ref{BLR_EF} summarizes results on efficiency in terms of relative improvement of MMHMC compared to HMC, measured in terms of time-normalized minimum ESS and maximum MCSE across variates, obtained using the best set of simulation parameters among the tested ones for each method. Based on these results we can conclude that for the BLR model and tested data sets, MMHMC demonstrates improvement over HMC of up to 2.5 times.

\subsubsection{Stochastic Volatility Model}

Stochastic volatility (SV) models are a useful tool for modeling time-varying volatility with significant potential for applications (e.g.\ risk management/risk prediction, pricing of financial derivatives). 

We consider the standard SV model defined with the latent, log-volatilities following autoregressive AR(1) process.
The model, as described by \citet{KSC98},  takes the following form 
\begin{eqnarray*}
	y_t &=&\beta \exp(x_t/2)\epsilon_t, \; \;  \;  \;  \epsilon_t \sim \mathcal{N}(0, 1)  \\
	x_{t} &=& \phi x_{t-1}  + \sigma \eta_t, \;  \;  \;  \; \; \:  \eta_t \sim \mathcal{N}(0,1) \\
	x_1 & \sim& \mathcal{N}\left(0,\frac{\sigma^2}{1-\phi^2}\right),
\end{eqnarray*}
where $y_t$ are observed data of mean corrected log-returns, equidistantly spaced in time for $t=1,\dots,T$, and $x_t$ are latent variables of log-volatility assumed to follow a stationary process. 
This assumption leads to the constraint $|\phi|<1$.
The error terms $\epsilon_t$ and $\eta_t$ are serially and mutually uncorrelated white noise sequences with the standard normal distribution.
The parameter $\beta$ of the model can be interpreted as the modal instantaneous volatility, $\phi$ as the persistence in the volatility and $\sigma$ as the volatility of the log-volatility, leading to the second constraint $\sigma>0$. 

Let denote the vector of model parameters as $\boldsymbol\theta=(\beta, \sigma, \phi)$. Its priors are chosen as
$p(\beta) \propto 1/\beta, 
\sigma^2 \sim \text{Scale-inv-}\chi^2(10,0.05), 
(\phi+1)/2 \sim \text{Beta}(20,1.5)$,
leading to 
\begin{eqnarray*} 
	p(\beta) &\propto& \frac 1 \beta \\
	p(\sigma) &\propto& \sigma^{-11} \exp\{ -1/4\sigma^2\} \\
	p(\phi)  &\propto& \left(\phi+1\right)^{19} \left(1-\phi \right)^{\frac{1}{2}}.
\end{eqnarray*}

Instead of sampling jointly model parameters and latent volatilities from $\pi(\boldsymbol\theta, \mathbf x| \mathbf y)$, we follow a common procedure of cycling through the two full conditional distributions $\pi(\boldsymbol\theta| \mathbf y, \mathbf x)$ and $\pi(\mathbf x| \mathbf y,\boldsymbol\theta)$ (see e.g.\ \cite{JPR94,Chen:2000,Liu08}).

Since HMC methods sample real valued parameters, we 
handle the constraints $\sigma^2>0$ and $-1 \leq \phi \leq 1$ by making use of the transformation  $\mathcal T: \boldsymbol\theta \rightarrow \bar{\boldsymbol\theta}$ to the real line, defined as
$$\bar{\boldsymbol\theta}=\mathcal T(\boldsymbol\theta)
=\left(\beta, \ln(\sigma), \text{artanh}(\phi) \right)=(\beta,\gamma,\alpha)$$
with the Jacobian
\begin{equation*}
\mathcal J_{\mathcal T}=\begin{bmatrix}
\frac{\mathrm d\beta}{\mathrm d\beta}& 0 & 0           \\[0.3em]
0 & \frac{\mathrm d\gamma}{\mathrm d\sigma}         & 0 \\[0.3em]
0           & 0 &\frac{\mathrm d\alpha}{\mathrm d\phi} 
\end{bmatrix}
=\begin{bmatrix}
1 & 0 & 0           \\[0.3em]
0 & \sigma^{-1}         & 0 \\[0.3em]
0           & 0 & (1-\phi^2)^{-1}
\end{bmatrix},
\end{equation*}
which accounts for the change of variables within the Hamiltonian dynamics and Metropolis test.

\paragraph{Experimental setting.} We examine sampling of the standard SV model on simulated data with values $\beta=0.65, \sigma=0.15, \phi=0.98$, for $T=2000,5000,10000$ time points. This results in three experiments of dimensions $D=2003,5003,10003$, which include three model parameters and $T$ latent volatility variables to sample. 
We run 10000 iterations as a warm-up and generate 100000 posterior samples  collecting every 5th sample. 
We compare MMHMC with HMC, and for $D=2003$ we additionally run the RMHMC and GSHMC methods. 
The simulation parameters of the four methods are summarized in Appendix \ref{Sec:Exp_setup}.
The results presented in this section for MMHMC are obtained with the M-ME3 and M-ME2 integrators for $D=2003$ and $D=5003,10003$, respectively, and the modified Hamiltonian \eqref{modHam4_an}. 
{\color{black}As proposed in the original paper, we run GSHMC with modified Hamiltonians calculated using numerical derivatives.
	However, we notice that the original implementation of derivatives in GSHMC is less efficient than the one in \textsf{HaiCS} and thus the GSHMC performance in the following comparison is likely overestimated.}

\begin{figure}[ht!]
	\centering
	{\includegraphics[width=0.8\columnwidth ]{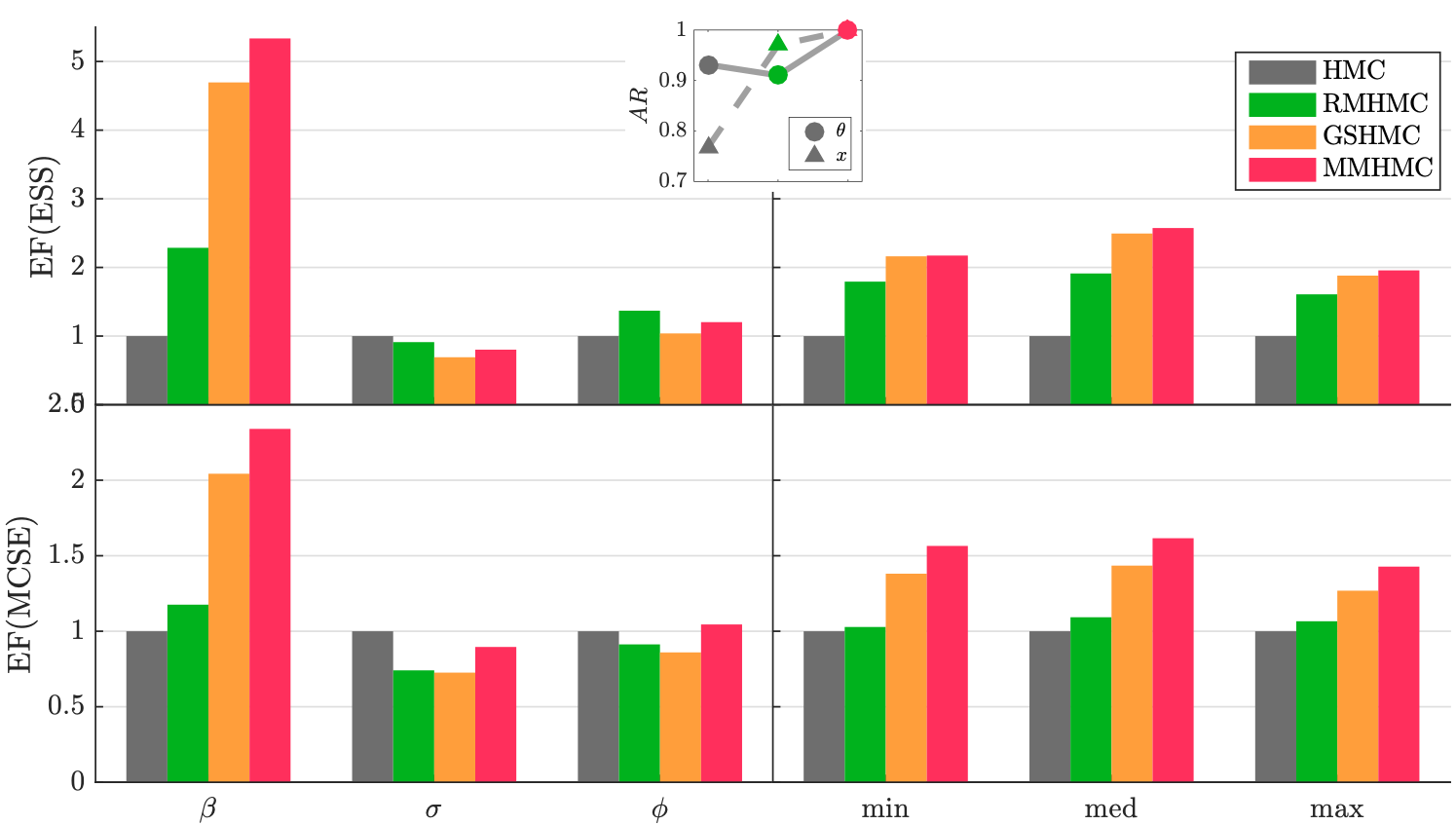} }
	\caption{Stochastic volatility. Sampling efficiency of RMHMC and MMHMC relative to HMC in terms of time-normalized ESS (top) and MCSE (bottom) 
		for SV model parameters (left) and latent variables (right) and corresponding acceptance rates (inset) for dimension $D=2003$.}
	\label{Fig:SV_2000}
\end{figure}

\begin{figure}[ht!]
	\centering
	{\includegraphics[width=0.8\columnwidth ]{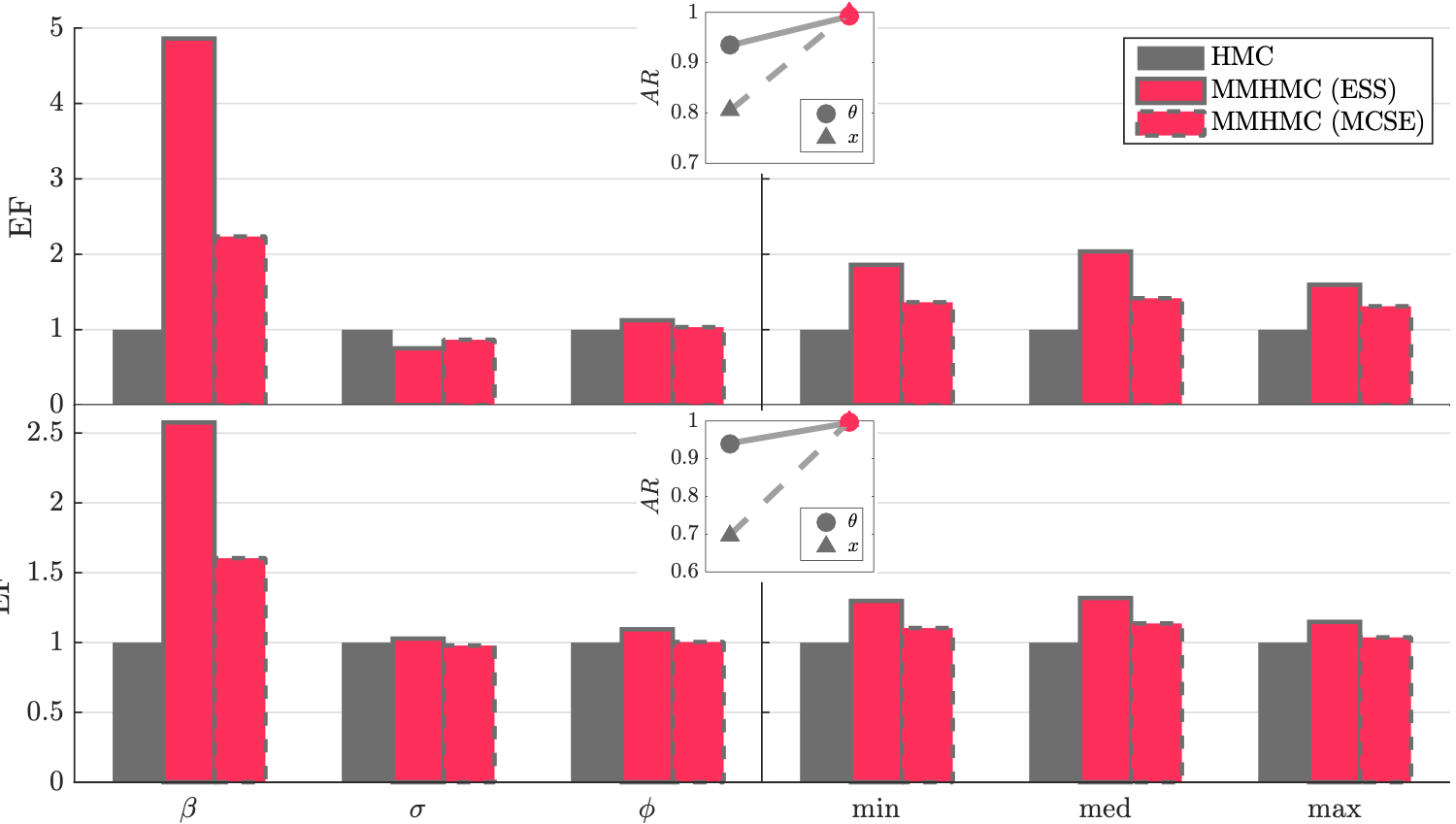} }
	\caption{Stochastic volatility. Sampling efficiency of MMHMC relative to HMC in terms of time-normalized ESS and MCSE for SV model parameters (left) and latent variables (right) and corresponding acceptance rates (inset) for dimensions $D=5003$ (top) and $D=10003$ (bottom).}
	\label{Fig:SV_5000}
\end{figure}

\paragraph{Results.} 
Figures \ref{Fig:SV_2000} and \ref{Fig:SV_5000} provide efficiency in terms of time-normalized ESS and MCSE 
relative to HMC for experiments with $D=2003$ and $D=5003,10003$, respectively. Acceptance rates (shown in inset figures) are rather high for all methods. However, there is no clear connection between obtained acceptance rates and ESS/MCSE.
{\color{black}Results for $D=2003$ demonstrate that RMHMC, GSHMC and MMHMC, outperform HMC in terms of time-normalized ESS for $\beta$ and latent variables. However, all tested methods sample $\sigma$ and $\phi$ comparably.
	For all sampled parameters, MMHMC shows comparable or superior performance to RMHMC.}

{\color{black}We recall here that in contrast to the RMHMC method, HMC and MMHMC use the  identity mass matrix. One way to improve the performance of these methods compared to RMHMC would be to define the mass matrix from an estimate of global covariances in the warm-up phase and use it for obtaining the posterior samples. }

We do not have an access to the optimal parameters for RMHMC for the  dimensions higher than $D=2003$. For $D=5003,10003$ we compare only MMHMC and HMC and observe that the superiority of MMHMC  for sampling of model parameters and latent variables is maintained for higher dimensions.

\section{Conclusions}\label{Sec:Conclusions}

We developed the irreversible MCMC method for enhanced statistical sampling, which offers higher sampling efficiency than the state-of-the-art MCMC method, Hamiltonian Monte Carlo.
Our new approach, called Mix \& Match HMC (MMHMC) arose as an extension of Generalized Shadow Hybrid Monte Carlo (GSHMC), earlier proposed for molecular simulation, published, patented and successfully tested on complex physical systems \cite{Akhmatskaya08,Wee08,Akhmatskaya:2009,ARN11,Escribano:2017,Bonilla:2018, FGarcia:2019}.
The MMHMC introduces a number of modifications in GSHMC needed for efficient sampling in computational statistics. It can be viewed as a generalized HMC 
importance sampler---momentum is updated in a general form and sampling is performed with respect to 
an importance distribution that is defined through modified Hamiltonian.
To the best of our knowledge, this is the first time that the method
sampling with modified Hamiltonians has been implemented and applied to
Bayesian inference problems in computational statistics.

Being a method that generates both correlated and weighted samples, MMHMC requires a metric for sampling efficiency different from the one commonly used for MCMC. Here we suggested such a metric suitable for MCMC importance sampling based methods.

The method has been carefully tested and compared with the traditional and advanced sampling techniques 
such as Random Walk  Metropolis-Hastings, Metropolis Adjusted Langevin Algorithm, Hamiltonian Monte Carlo, Riemann Manifold Hamiltonian Monte Carlo, Generalized Hybrid Monte Carlo and Generalized Shadow Hybrid Monte Carlo.

When compared to HMC, RWMH, MALA, GHMC and GSHMC, the MMHMC method demonstrates superior performance, in terms of higher acceptance rate, bigger time-normalized ESS and smaller MCSE, for a range of applications, range of dimensions and choice of  parameters of the methods. 
The improvements are bigger for high-dimensional problems---for the multivariate Gaussian problem MMHMC demonstrated an improvement over HMC of up to 29 times. When comparing only for the best set of parameters among the tested ones for each method, MMHMC shows around 17 times better performance than HMC for the Gaussian problem and around 2.5 times improvement for the BLR model.

MMHMC and RMHMC demonstrate comparable, with a slight advantage of MMHMC, performance for the tested SV model. However, in contrast to the original RMHMC, MMHMC  does not rely on higher order derivatives or inverse of the metric, and thus requires less implementation and computational effort. This issue becomes particularly important for high-dimensional problems with dense Hessian matrix. In addition, choices of integrators for RMHMC are limited due to the use of non-separable Hamiltonians, whereas MMHMC is well compatible with advanced splitting integration schemes.

\vspace{0.7cm}

{\small 
\textbf{Acknowledgments.}
The authors would like to thank the financial support from MTM2016-76329-R (AEI/FEDER, EU) funded by the Spanish Ministry of Economy and Competitiveness (MINECO) and from Basque Government - ELKARTEK Programme, grant KK-2018/00054.
This work has been possible thanks to the support of the computing infrastructure of the i2BASQUE academic network, and the technical and human support provided by IZO-SGI SGIker of UPV/EHU and European funding	(ERDF and ESF). 
This research is also supported by the Basque Government through the BERC 2018-2021  program and by MINECO: BCAM Severo Ochoa accreditation SEV-2017-0718.

This work was also part of the Agile BioFoundry (http://agilebiofoundry.org) supported by the U.S. Department of Energy, Energy Efficiency and Renewable Energy, Bioenergy Technologies Office, through contract DE-AC02-05CH11231 between Lawrence Berkeley National Laboratory and the U.S. Department of Energy. The views and opinions of the authors expressed herein do not necessarily state or reflect those of the United States Government or any agency thereof.
Neither the United States Government nor any agency thereof, nor any of their employees, makes any warranty, expressed or implied, or assumes any legal liability or responsibility for the accuracy, completeness, or usefulness of any information, apparatus, product, or process disclosed, or represents that its use would not infringe privately owned rights. 
The United States Government retains and the publisher, by accepting the article for publication, acknowledges that the United States Government retains a nonexclusive, paid-up, irrevocable, worldwide license to publish or reproduce the published form of this manuscript, or allow others to do so, for United States Government purposes. 
The Department of Energy will provide public access to these results of federally sponsored research in accordance with the DOE Public Access Plan (http://energy.gov/downloads/doe-public-access-plan).

\appendix

\section{Invariance of the PMMC step}\label{App:Invariance}

The Partial Momentum Monte Carlo step of the MMHMC method leaves the importance target distribution $\tilde{\pi}$ (Eq.\ \ref{modDistrMMHMC}) invariant if for a transition kernel $T(\cdot|\cdot)$ the following condition is satisfied
\begin{equation*}
\tilde{\pi}(\boldsymbol{\theta}',\mathbf{p}')=\int \tilde{\pi}(\boldsymbol{\theta},\mathbf{p})T\left((\boldsymbol{\theta}',\mathbf{p}')|(\boldsymbol{\theta},\mathbf{p})\right) \mathrm{d}\boldsymbol{\theta} \mathrm{d}\mathbf{p}
\end{equation*} 
for all $n=1,\dots,N$. 

The PMMC step is sampling on a space augmented with a noise vector $\mathbf u \sim \mathcal{N}(0,M)$ with the extended density $\hat{\pi}$ (defined in Eq.\ \ref{extended_target_prob}), for which
$$\tilde{\pi}(\boldsymbol{\theta},\mathbf{p})=\int \hat{\pi}(\boldsymbol{\theta},\mathbf{p},\mathbf{u})\mathrm{d}\mathbf{u} .$$
Therefore, we want to show that
\begin{equation}\label{eq:invariance_ext}
\hat{\pi}(\boldsymbol{\theta}',\mathbf{p}',\mathbf{u}')=\int \hat{\pi}(\boldsymbol{\theta},\mathbf{p},\mathbf{u})T\left((\boldsymbol{\theta}',\mathbf{p}',\mathbf{u}')|(\boldsymbol{\theta},\mathbf{p},\mathbf{u})\right) \mathrm{d}\boldsymbol{\theta} \mathrm{d}\mathbf{p} \mathrm{d}\mathbf{u},
\end{equation}
for the transition kernel defined as
\begin{equation*}
\begin{aligned}
T\left((\boldsymbol{\theta}',\mathbf{p}',\mathbf{u}')|(\boldsymbol{\theta},\mathbf{p},\mathbf{u})\right)&=\mathcal P\cdot \delta \left((\boldsymbol{\theta}',\mathbf{p}',\mathbf{u}')-\mathcal R(\boldsymbol{\theta},\mathbf{p},\mathbf{u}) \right)\\
&+(1-\mathcal P)\cdot\delta \left((\boldsymbol{\theta}',\mathbf{p}',\mathbf{u}')-\hat{\mathcal F}(\boldsymbol{\theta},\mathbf{p},\mathbf{u})\right),
\end{aligned}
\end{equation*}
where $\mathcal P=\min\left\{1,\hat{\pi}(\mathcal{R}(\boldsymbol{\theta},\mathbf{p},\mathbf{u}))/\hat{\pi}(\boldsymbol{\theta},\mathbf{p},\mathbf{u})\right\}$ is the Metropolis probability, $\delta$ is the Delta function, $\mathcal R$ is the proposal function (Eq.\ \ref{eq:PMMCorig}) and $\hat{\mathcal F}(\boldsymbol{\theta},\mathbf{p},\mathbf{u})=(\boldsymbol{\theta},\mathbf{p},-\mathbf{u})$ is the flipping function. Note that the map $\mathcal R$ is volume preserving, hence, the Metropolis probability $\mathcal P$ does not inlcude the Jacobian factor. For the sake of clarity, we denote $\mathbf x=(\boldsymbol{\theta},\mathbf{p},\mathbf{u})$ and write the right-hand side of the expression \eqref{eq:invariance_ext} as
\begin{equation*}
\begin{aligned}
\int T(\mathbf x'|\mathbf x)\hat{\pi}(\mathbf x) \mathrm{d}\mathbf x&=\underbrace{\int \min\left\{\hat{\pi}(\mathbf x),\hat{\pi}(\mathcal R(\mathbf x)) \right\} \cdot \delta\left(\mathbf x'- \mathcal R(\mathbf x) \right)\mathrm{d}\mathbf x}_\text{1st term}\\
&\underbrace{+\int \hat{\pi}(\mathbf x)  \cdot \delta\left(\mathbf x'- \hat{\mathcal F}(\mathbf x) \right)\mathrm{d}\mathbf x}_\text{2nd term} \\
&\underbrace{-\int \min\left\{\hat{\pi}(\mathbf x),\hat{\pi}(\mathcal R(\mathbf x)) \right\} \cdot \delta\left(\mathbf x'- \hat{\mathcal F}(\mathbf x) \right)\mathrm{d}\mathbf x}_\text{3rd term}.
\end{aligned}
\end{equation*}
Applying change of variables $\mathbf x=\hat{\mathcal F} \circ \mathcal R(\bar{\mathbf x})$, which is volume preserving, to the 1st term in the sum, omitting the bars, and using the fact that $\hat{\mathcal F} = \mathcal R\circ\hat{\mathcal F}\circ\mathcal R$, one obtains 
$$\text{1st term}=\int \min\left\{\hat{\pi}(\hat{\mathcal F} \circ\mathcal R(\mathbf x)),\hat{\pi}(\hat{\mathcal F} (\mathbf x)) \right\} \cdot \delta\left(\mathbf x'- \hat{\mathcal F}(\mathbf x) \right)\mathrm{d}\mathbf x.$$
Since $\hat\pi\circ\hat{\mathcal F}=\hat{\pi}$, the 1st and 3rd terms cancel out. Employing change of variables $\mathbf x=\hat{\mathcal F}(\bar{\mathbf x})$ to the 2nd term and again omitting the bars, leads to
$$\text{2nd term}=\int \hat{\pi}(\mathbf x)  \cdot \delta\left(\mathbf x'- \mathbf x \right)\mathrm{d}\mathbf x=\hat{\pi}(\mathbf x'),$$
which proves the equality \eqref{eq:invariance_ext}.

\section{Modified Hamiltonians for Splitting Integrators}\label{App:ModHam}
The coefficients for the two-stage integrator family \eqref{2S} and modified Hamiltonians \eqref{modHam4_an}--\eqref{modHam6_an_Gauss} are the following
{\small 
	\begin{equation}\label{modHam2S}
	\begin{aligned}
	c_{21}=& \frac{1}{24}\Big(6b-1\Big) \\
	c_{22}=& \frac{1}{12}\Big(6b^2-6b+1\Big)\\
	c_{41} =& \frac{1}{5760}\Big(7-30b\Big) \\
	c_{42} =& \frac{1}{240}\Big(-10b^2+15b-3\Big) \\
	c_{43} =& \frac{1}{120}\Big(-30b^3+35b^2-15b+2\Big) \\
	c_{44} =& \frac{1}{240}(20b^2-1).
	\end{aligned}
	\end{equation}
}

Using \eqref{modHam2S} one can also obtain the modified Hamiltonian for the Verlet integrator, since two steps of Verlet integration are equivalent to one step of the two-stage integrator with $b=1/4$. The coefficients are therefore
\begin{eqnarray*}\label{modHamVerlet}
	c_{21}&=& \frac{1}{12}, \hspace{0.5cm} c_{22}=- \frac{1}{24} \\
	c_{41}& = &-\frac{1}{720}, \hspace{0.5cm} c_{42} = \frac{1}{120}, \hspace{0.5cm}c_{43} = -\frac{1}{240}, \hspace{0.5cm} c_{44} =\frac{1}{60}. \nonumber
\end{eqnarray*}

For three-stage integrators \eqref{3S} (a two-parameter family) the coefficients are
\begin{equation*}\label{modHam3S}
\begin{aligned}
c_{21}=& \frac{1}{12} \Big(1-6a(1-a)(1-2b)\Big)\\
c_{22}=& \frac{1}{24} \Big(6a(1-2b)^2-1\Big)\\
c_{41} =& \frac{1}{720}\Big( 1 + 2 (a-1) a (8 + 31 (a-1) a) (1 - 2 b) - 4 b\Big)\\
c_{42}=& \frac{1}{240} \Big(6 a^3 (1 - 2 b)^2 - 
a^2 (19 - 116 b + 36 b^2 + 240 b^3)\\
&+a (27 - 208 b + 308 b^2)- 48 b^2+ 48 b    -7\Big) \\
c_{43} =&\frac{1}{180}\Big(1 + 15 a (1- 2 b) (-1 + 2 a (2 - 3 b + a (4 b-2)))\Big)  \\
c_{44} =& \frac{1}{240} \Big(-1 + 20 a (1 - 2 b) (b + a (1 + 6 (b-1) b))\Big).
\end{aligned}
\end{equation*}

The coefficients for the modified Hamiltonians \eqref{modHam4num}--\eqref{modHam6num} are calculated as
\begin{eqnarray*}\label{k_coeff}
	k_{21}&=&c_{21}, \hspace{0.3cm} k_{22}=c_{22}, \\
	k_{41}&=&c_{41}, \hspace{0.3cm} k_{42}=3c_{41} +c_{42}, \\
	k_{43}&=&c_{41} +c_{44}, \hspace{0.3cm}  k_{44}=3c_{41}+c_{42}+c_{43}.\nonumber
\end{eqnarray*} 

For the 4th order modified Hamiltonian  \eqref{modHam4num} we use the second order centered finite difference approximations of time derivatives of the gradient of the potential function\begin{equation}\label{timeder_modH4}
\mathbf U^{(1)}=\frac{\mathbf U(t_{n+1})-\mathbf U(t_{n-1})}{2\varepsilon},
\end{equation}
with $\varepsilon=h$ for the Verlet, $\varepsilon=h/2$ for two-stage and $\varepsilon=ah$ for three-stage integrators with $a$ being  the integrator's coefficient advancing position variables.
The 6th order modified Hamiltonian \eqref{modHam6num}, here considered only for the Verlet and two-stage integrators, is calculated using 
fourth order approximation 
for the first derivative and second order approximations for the second and third derivatives
\begin{eqnarray*}
	\mathbf U^{(1)}&= &\frac{\mathbf U(t_{n-2})-8\mathbf U(t_{n-1})+8\mathbf U(t_{n+1})-\mathbf U(t_{n+2})}{12\varepsilon} \nonumber \\
	\mathbf U^{(2)}&= &\frac{\mathbf U(t_{n-1})-2\mathbf U(t_n)+\mathbf U(t_{n+1})}{\varepsilon^2}\\
	\mathbf U^{(3)}&=&\frac{-\mathbf U(t_{n-2})+2\mathbf U(t_{n-1})-2\mathbf U(t_{n+1})+\mathbf U(t_{n+2})}{2\varepsilon^3},\nonumber
\end{eqnarray*}
where $\varepsilon$ depends on the integrator as before.  
The interpolating polynomial in terms of the gradient of the potential function $\mathbf U(t_i)= U_{\boldsymbol{\theta}}(\boldsymbol{\theta}^i), \; i=n-k,\dots,n,\dots,n+k, n \in \{0, L\}$ is constructed from a numerical trajectory $\{U_{\boldsymbol{\theta}}(\boldsymbol{\theta}^i\}^{L+k}_{i=-k}$ where $k=1$ and $k=2$ for the 4th and 6th order modified Hamiltonians, respectively.

\section{Modified PMMC Step}\label{App:ModPMMC}

In the modified PMMC step proposed for MMHMC, a partial momentum update is integrated into the modified Metropolis test, i.e.\ it is implicitly present in the algorithm.  
This reduces the frequency of derivative calculations in the Metropolis function.   
To implement this idea, one should recall that the momentum update probability 
\begin{equation}\label{PMMC_prob}
\mathcal{P}=\min\left\{1,\frac{\exp\big(-\hat{H}(\boldsymbol{\theta},\mathbf p^*,\mathbf u^*)}{\exp\big(-\hat{H}(\boldsymbol{\theta},\mathbf p,\mathbf u)\big)}\right\}
\end{equation}
depends on the error in the extended Hamiltonian \eqref{extendedHam}. 
Let us first consider the 4th order modified Hamiltonian  \eqref{modHam4_an} with analytical derivatives of the potential function. 
It is easy to show that the difference in the extended Hamiltonian \eqref{extendedHam} between a current state and a state with partially updated momentum is

\begin{align}\label{deltaHam4an_PMMC}
\Delta{\hat H} 
&= {U(\boldsymbol{\theta})} + \frac{1}{2} (\mathbf p^*)^T M^{-1} \mathbf p^*\nonumber \\ 
&+ h^2c_{21}(\mathbf p^*)^T M^{-1}U_{\boldsymbol{\theta} \boldsymbol{\theta} }(\boldsymbol{\theta} ) M^{-1} \mathbf p^*\nonumber \\ 
&+  {h^2c_{22}U_{\boldsymbol{\theta} }(\boldsymbol{\theta} ) M^{-1}U_{\boldsymbol{\theta}}(\boldsymbol{\theta})} +  \frac{1}{2} (\mathbf u^*)^T M^{-1}\mathbf u^* \nonumber \\ 
&- {U(\boldsymbol{\theta})}  - \frac{1}{2} \mathbf p^T M^{-1} \mathbf p - h^2c_{21}\mathbf p^T M^{-1}U_{\boldsymbol{\theta} \boldsymbol{\theta} }(\boldsymbol{\theta} ) M^{-1} \mathbf p \nonumber \\ 
&-  {h^2c_{22}U_{\boldsymbol{\theta}}(\boldsymbol{\theta}) M^{-1}U_{\boldsymbol{\theta}}(\boldsymbol{\theta})} - \frac{1}{2} \mathbf u^T M^{-1} \mathbf u \nonumber \\
&= {h^2}c_{21}\Big(\varphi A + 2\sqrt{\varphi(1-\varphi)}B \Big)
\end{align}
with
\begin{equation}\label{AB_deltaHam4an_PMMC}
\begin{aligned}
A&=(\mathbf u-\mathbf p)^T U_{\boldsymbol{\theta} \boldsymbol{\theta} }(\boldsymbol{\theta} ) (\mathbf u+\mathbf p) \\
B&=\mathbf {u}^TU_{\boldsymbol{\theta} \boldsymbol{\theta} }(\boldsymbol{\theta} ) \mathbf p.
\end{aligned}
\end{equation}
For the 6th order modified Hamiltonian \eqref{modHam6_an_Gauss} for Gaussian problems, the error in the extended Hamiltonian  \eqref{extendedHam} 
can be calculated in a similar manner 
\begin{equation}\label{deltaHam6anGauss_PMMC}
\begin{aligned}
\Delta{\hat H}  &= {h^2}c_{21}\Big(\varphi (A-B) + 2\sqrt{\varphi(1-\varphi)}C \Big)\\
&+{h^4}c_{44}\Big(\varphi (D-E) + 2\sqrt{\varphi(1-\varphi)}F \Big),
\end{aligned}
\end{equation}
with
\begin{equation*}
\begin{aligned}
A&=\mathbf u^T U_{\boldsymbol{\theta} \boldsymbol{\theta} }(\boldsymbol{\theta} )\mathbf u \\
B&=\mathbf {p}^TU_{\boldsymbol{\theta} \boldsymbol{\theta} }(\boldsymbol{\theta} ) \mathbf p\\
C&=\mathbf {u}^TU_{\boldsymbol{\theta} \boldsymbol{\theta} }(\boldsymbol{\theta} ) \mathbf p\\
D&=(U_{\boldsymbol{\theta} \boldsymbol{\theta} }(\boldsymbol{\theta} ) \mathbf u)^TU_{\boldsymbol{\theta} \boldsymbol{\theta} }(\boldsymbol{\theta} ) \mathbf u\\
E&=(U_{\boldsymbol{\theta} \boldsymbol{\theta} }(\boldsymbol{\theta} ) \mathbf p)^TU_{\boldsymbol{\theta} \boldsymbol{\theta} }(\boldsymbol{\theta} ) \mathbf p\\
F&=(U_{\boldsymbol{\theta} \boldsymbol{\theta} }(\boldsymbol{\theta} ) \mathbf u)^TU_{\boldsymbol{\theta} \boldsymbol{\theta} }(\boldsymbol{\theta} ) \mathbf p.\\
\end{aligned}
\end{equation*}
Therefore, if the modified Hamiltonians  \eqref{modHam4_an}--\eqref{modHam6_an_Gauss} with analytical derivatives are used, a new momentum can be determined as
\begin{equation}\label{newPMMC_Metropolis}
\bar{\mathbf p}=\left\{\begin{array}{l l}
\sqrt{1-{\varphi}}\mathbf p+\sqrt{{\varphi}}\mathbf u  & \mbox{ with probability } \\
& \hspace{0.5cm} \mathcal{P}= \min\{1,\exp(-\Delta\hat{H})\}\\
\mathbf p & \mbox{ otherwise,}
\end{array}\right.
\end{equation}
where $\mathbf u \sim \mathcal{N}(0,M)$ is the noise vector, $\varphi \in \left(0,1\right]$ and $\Delta \hat H$ is defined as in \eqref{deltaHam4an_PMMC} or \eqref{deltaHam6anGauss_PMMC}.

Consequently, for models with no hierarchical structure, there is no need to calculate gradients within the PMMC step, second derivatives can be taken from the previous  Metropolis test within the HDMC step, and there is no need to generate $\mathbf u^*$.

If the modified Hamiltonians are calculated using numerical time derivatives of the gradient of the potential function, for the Verlet, two- and three-stage integrators as in \eqref{modHam4num}--\eqref{modHam6num}, the difference in the  4th order extended Hamiltonian becomes
\begin{equation}\label{deltaHam4num_PMMC}
\Delta{\hat H}= {h}k_{21}\Big((\mathbf p^*)^T P^*_1 -\mathbf p^TP_1 \Big),
\end{equation}
whereas for the 6th order extended Hamiltonian it is
\begin{equation*}
\begin{aligned}
\Delta{\hat H}&=hk_{21}\Big((\mathbf p^*)^T P^*_1 -\mathbf p^TP_1 \Big)\\
&+hk_{41}\Big((\mathbf p^*)^T P^*_3 -\mathbf p^TP_3 \Big)\\ 
&+h^2k_{42}\Big(U_{\mathbf x}^T P^*_2 - U_{\mathbf x}^TP_2 \Big)\\
&+h^2k_{43}\Big((P^*_1)^T P^*_1 -P_1^TP_1 \Big).
\end{aligned}
\end{equation*}
Here $P^*_1, P^*_2,P^*_3$, are the first, second and third order scaled time derivatives of the gradient, respectively (see Section \ref{Sec:ModHam}), calculated from the trajectory with updated momentum $\mathbf p^*$.
The computational gain of the new PMMC step, in this case, results from skipping a calculation of the terms multiplying $k_{22}$ in \eqref{modHam4num} and $k_{44}$ in \eqref{modHam6num}. 
It has to be admitted that the term multiplying $k_{22}$ in \eqref{modHam4num} is of negligible cost, and thus the gain from using the new momentum update is not as significant as in the case of modified Hamiltonians with analytical derivatives. On the contrary, the saving in computation arising from the absence of the term multiplying $k_{44}$ in the 6th order modified Hamiltonian \eqref{modHam6num}, is essential. 

In summary, in the case of the 6th order modified Hamiltonian, with derivatives calculated either analytically or numerically, the proposed momentum refreshment enhances computational performance of MMHMC. This also applies to the cases when the 4th order modified Hamiltonian with analytical derivatives is used. In this situation, however, if the Hessian matrix of the potential function is dense, instead of using the modified Hamiltonian with analytical derivatives, we recommend using numerical derivatives, for which the saving is negligible. On the other hand, if the computation of the Hessian matrix is not very costly (e.g.\ being block-diagonal, sparse, close to constant), it might be more efficient to use analytical derivatives, for which the new formulation of the Metropolis test leads to computational saving.

\section{Algorithmic Summary}\label{Sec:Algorithm}

\begin{algorithm}
	\caption{Hamiltonian Monte Carlo}
	\label{AlgHMC}
	\begin{algorithmic}[1]
		\State \textbf{Input:} $N$: number of Monte Carlo samples
		\newline
		{\color{white}\textbf{Input:}} $h$: step size
		\newline {\color{white}\textbf{Input:}} $L$: number of integration steps
		\newline {\color{white}\textbf{Input:}} $M$: mass matrix
		\newline {\color{white}\textbf{Input:}} $\Psi_{h,L}$: numerical integrator
		\STATE Initialize $\boldsymbol{\theta}^0$
		\FOR{$n=1,\dots, N$}
		\State $\boldsymbol{\theta}=\boldsymbol{\theta}^{n-1}$
		\State Draw momentum from Gaussian distribution: $\mathbf p\sim \mathcal N (0,M)$
		\State Generate a proposal by integrating Hamiltonian dynamics: $(\boldsymbol{\theta}',\mathbf p')= \Psi_{h,L}(\boldsymbol{\theta},\mathbf p)$
		\State Set $\boldsymbol{\theta}^n=\boldsymbol{\theta}'$ with probability $\alpha=\min\{1,\exp(H(\boldsymbol{\theta},\mathbf p)-H(\boldsymbol{\theta}',\mathbf p'))\}$,
		otherwise set $\boldsymbol{\theta}^n=\boldsymbol{\theta}$
		\State Discard momentum $\mathbf p'$
		\ENDFOR
	\end{algorithmic}
\end{algorithm}

We provide two alternative algorithms for the MMHMC method. One (Algorithm \ref{AlgMMHMC_an}) uses the modified Hamiltonians defined through analytical derivatives of the potential function and is recommended for the problems with sparse Hessian matrices. The other algorithm (Algorithm \ref{AlgMMHMC_num}) relies on the modified Hamiltonians expressed through numerical time derivatives of the gradient of the potential function. This algorithm, although including additional integration step, is beneficial for cases where higher order derivatives are computationally demanding. 

\begin{algorithm} 
	\caption{MMHMC using Hessian of the potential function}
	\label{AlgMMHMC_an}
	\begin{algorithmic}[1]
		\State \textbf{Input:} $N$: number of Monte Carlo samples
		\newline\hspace*{1cm} $h$: step size   
		\newline \hspace*{1cm} $p(L)$:  number-of-integration-steps randomization policy
		\newline \hspace*{1cm} $p(\varphi)$: noise-parameter randomization policy
		\newline \hspace*{1cm} $M$: mass matrix
		\newline \hspace*{1cm} $r$: number of stages in the  numerical integrator ($r=1,2,3$)
		\newline \hspace*{1cm} $\Psi_{h,L}$: symplectic $r$-stage numerical integrator
		\State Initialize $(\boldsymbol\theta^0,\mathbf p^0)$
		\State Calculate Hessian $U_{\boldsymbol\theta\boldsymbol\theta}(\boldsymbol\theta^0)$
		\FOR{$n=1,\dots, N$}
		\State Draw  $L_n\sim p(L), \varphi_n\sim p(\varphi)$ 
		\State $(\boldsymbol\theta,\mathbf p)=(\boldsymbol\theta^{n-1},\mathbf p^{n-1})$
		\newline  \textcolor{white}{xx}PMMC step 
		\State Draw noise $\mathbf u \sim \mathcal{N}(0, M)$ and update momenta
		\begin{equation*}
		\bar{\mathbf p}=\left\{\begin{array}{l l}
		\sqrt{1-{\varphi_n}}\mathbf p+\sqrt{{\varphi_n}}\mathbf u  & \mbox{with probability } \\
		& \hspace{0.5cm} \mathcal{P}= \min\left\{1,\exp\left(-\Delta\hat{H}\right)\right\} \\
		\mathbf p & \mbox{otherwise}
		\end{array}\right.
		\end{equation*}
		$\Delta\hat{H}$ defined in Eqs.\ \eqref{deltaHam4an_PMMC}--\eqref{AB_deltaHam4an_PMMC}
		\State Calculate modified Hamiltonian  $\tilde{H}^{[4]}(\boldsymbol\theta, \bar{\mathbf p})$ defined in Eq.\ \eqref{modHam4_an}
		\newline  \textcolor{white}{xx}HDMC step
		\State Generate a proposal by integrating Hamiltonian dynamics with step size $h$ over $L_n$ steps 
		$$(\boldsymbol\theta',\mathbf p')= \Psi_{h,L_n}(\boldsymbol\theta, \bar{\mathbf p} )$$           
		\State Calculate Hessian $U_{\boldsymbol\theta\boldsymbol\theta}(\boldsymbol\theta')$ and modified Hamiltonian $\tilde{H}^{[4]}(\boldsymbol\theta', \mathbf p')$
		\State Metropolis test
		\begin{equation*}
		( \boldsymbol\theta^{n}, \mathbf p^{n})=\left\{
		\begin{array}{l l}
		(\boldsymbol\theta', \mathbf p') & \quad  \text{ accept with probability } \\ & \hspace{0.5cm}\alpha=\min\left\{1,\exp\left(-\Delta\tilde{H}\right)\right\} \\
		(\boldsymbol\theta, -\mathbf p) & \quad \text{ reject otherwise} \\ 
		\end{array} 
		\right.
		\end{equation*}
		$\Delta\tilde{H}=\tilde{H}^{[4]}(\boldsymbol\theta',\mathbf p')-\tilde{H}^{[4]}(\boldsymbol\theta,\bar{\mathbf p})$  
		\ENDFOR
		\State Calculate weights $w_n,n=1,\dots,N$ (Eq.\ \ref{Eq:Weights})
		and estimate integral \eqref{Integral} as 
		$$\hat I = \frac{\sum_{n=1}^N f(\boldsymbol\theta^n) w_n }{\sum_{n=1}^N w_n} $$  
	\end{algorithmic}
\end{algorithm} 

\begin{algorithm}
	\caption{MMHMC using numerical derivatives of the gradient of the potential}
	\label{AlgMMHMC_num}
	\begin{algorithmic}[1]
		\State \textbf{Input:} $N$: number of Monte Carlo samples
		\newline\hspace*{1.12cm} $h$: step size   
		\newline \hspace*{1.12cm} $p(L)$:  number-of-integration-steps randomization policy
		\newline \hspace*{1.12cm} $p(\varphi)$: noise-parameter randomization policy
		\newline \hspace*{1.12cm} $M$: mass matrix
		\newline \hspace*{1.12cm} $r$: number of stages in the  numerical integrator ($r=1,2,3$)
		\newline \hspace*{1.12cm} $\Psi_{h,L}$: symplectic $r$-stage numerical integrator
		\State Initialize $(\boldsymbol\theta^0,\mathbf p^0)$
		\State Integrate one stage (i.e.\ one gradient calculation) backward, $\Psi_{h,-1}(\boldsymbol\theta^0, \mathbf p^0)$, and forward, $\Psi_{h,1}(\boldsymbol\theta^0, \mathbf p^0)$
		\State Calculate scaled time derivative of the gradient $P_1$ using Eq.\ \eqref{timeder_modH4}
		\FOR{$n=1,\dots, N$}
		\State Draw  $L_n\sim p(L), \varphi_n\sim p(\varphi)$ 
		\State $(\boldsymbol\theta,\mathbf p)=(\boldsymbol\theta^{n-1},\mathbf p^{n-1})$
		\newline  \textcolor{white}{xx}PMMC step 
		\State Draw noise $\mathbf u \sim \mathcal{N}(0, M)$ and propose momenta 
		$$\mathbf p^*=\sqrt{1-{\varphi_n}}\mathbf p+\sqrt{{\varphi_n}}\mathbf u$$
		\State Integrate one stage backward, $\Psi_{h,-1}(\boldsymbol\theta, \mathbf p^*)$, and forward,  $\Psi_{h,1}(\boldsymbol\theta, \mathbf p^*)$
		\State Calculate the resulting scaled time derivative of the gradient $P_1^*$ 
		\State Update momenta
		\begin{equation*}
		\bar{\mathbf p}=\left\{\begin{array}{l l}
		\mathbf p^*  & \mbox{ with probability }  \mathcal{P}= \min\{1,\exp(-\Delta\hat{H})\}, \\
		& \hspace{0.5cm}\Delta\hat{H} \text{ defined in Eq.\  \eqref{deltaHam4num_PMMC}}\\
		\mathbf p & \mbox{ otherwise}
		\end{array}\right.
		\end{equation*}
		
		\State Calculate modified Hamiltonian  $\tilde{H}^{[4]}(\boldsymbol\theta, \bar{\mathbf p})$ defined in Eq. \eqref{modHam4num}
		\newline \textcolor{white}{xx}HDMC step
		\State Integrate Hamiltonian dynamics with step size $h$ over $L_n^+$ steps and assign a proposal  \Comment{{\small $^+$ stands for an additional forward integration}}
		$$(\boldsymbol\theta',\mathbf p')= \Psi_{h,L_n}(\boldsymbol\theta, \bar{\mathbf p} )$$
		\State Calculate the resulting scaled time derivative of the gradient $P'_1$      
		\State Calculate modified Hamiltonian $\tilde{H}^{[4]}(\boldsymbol\theta', \mathbf p')$
		\State Metropolis test \Comment{{\small as in Algorithm \ref{AlgMMHMC_an}, line 11}}
		\ENDFOR
		\State  Compute weights and estimate integral \eqref{Integral} \Comment{{\small as in Algorithm \ref{AlgMMHMC_an}, line 14}}
	\end{algorithmic}
\end{algorithm} 

\section{Experimental setup}\label{Sec:Exp_setup}

\setlength{\tabcolsep}{6pt}
\renewcommand{\arraystretch}{1.1}

\begin{table*}[]
	\caption{Parameter values used for the multivariate Gaussian model experiments. The Verlet integrator was employed in HMC, GHMC and GSHMC methods, whereas for MMHMC the M-BCSS3 integrator was used for $D=100$ and M-ME3 for $D=1000,2000$. For HMC and GHMC step size is drawn from $U(0.8 h,1.2 h)$. For HMC, GHMC and MMHMC trajectory length is drawn from $U\{1,\dots,L \}$. For GHMC, MMHMC noise parameter is drawn from $U(0,\varphi)$. For GSHMC all parameters are fixed.}
	\centering
	\begin{tabular}{c c r ccccccc}
		\toprule
		$D$ &  {Method}  & &\multicolumn{7}{c}{Parameter value}   \\ 
		\midrule
		&   &  $h$ & 0.02&	0.03&	0.04&	0.05&	0.06&	0.07&	0.08	 \\ \cline{2-10}
		\multirow{8}{*}{100} & HMC &$L$& 500&	500&	500&500&500&500&400 \\ \cline{2-10}
		& \multirow{2}{*}{GHMC}  &$L$& 500&	500&	500&500&500&500&400 \\
		&& $\varphi$ & 0.1&	0.1&	0.1&	0.1&	0.9&	0.9&	0.9 \\ \cline{2-10}
		& \multirow{2}{*}{GSHMC} &$L$&	150&	100&	100&	100&	100&	100&	100\\
		&&$\varphi$	&0.1&	0.1&	0.1&	0.1&	0.1&	0.1&	0.1 \\ \cline{2-10}
		&\multirow{3}{*}{MMHMC}& $h$ & 0.06&	0.09&	0.12&	0.15&	0.18&	0.21&	0.24 \\
		& & $L$&	100&	67&	67&	67&	67&	67&	67\\ 
		&&$\varphi$	&0.1&	0.1&	0.1&	0.1&	0.1&	0.1&	0.1 \\
		\midrule
		\multirow{9}{*}{1000} &  & $h$& 0.006&	0.007&	0.008&	0.009&	0.01&	0.011&	0.012\\ \cline{2-10}
		& HMC &$L$& 5000 &	5000 &	5000 &	5000 &	5000 &	5000 &	5000  \\ \cline{2-10}
		& \multirow{2}{*}{GHMC}& $L$& 5000 &	5000 &	5000 &	5000 &	5000 &	5000 &	5000  \\
		&& $\varphi$ & 0.1&	0.1&	0.1&	0.1&	0.1&	0.1&	0.1 \\ \cline{2-10}
		& \multirow{2}{*}{GSHMC} &$L$&	2000&	1500&	1000&	1000&	1000&	1000&	1000\\
		&&$\varphi$	&0.1&	0.1&	0.1&	0.1&	0.1&	0.1&	0.1 \\ \cline{2-10}
		&\multirow{3}{*}{MMHMC}& $h$ & 0.018&	0.021&	0.024&	0.027&	0.03&	0.033&	0.036 \\
		& & $L$&	1333&	1000&667&	667&	667&	667&	667 \\ 
		&&$\varphi$	&0.1&	0.1&	0.1&	0.1&	0.1&	0.1&	0.1 \\
		\midrule
		\multirow{9}{*}{2000} &  & $h$& 0.003&	0.004&	0.005&	0.006&	0.007&	0.008\\ \cline{2-9}
		& HMC &$L$& 10000 &	10000 &	10000 &	10000 &	10000 &	10000 &	  \\ \cline{2-9}
		& \multirow{2}{*}{GHMC}& $L$& 10000 &	10000 &	10000 &	10000 &	10000 &	10000 &	  \\ 
		&& $\varphi$ & 0.1&	0.1&	0.1&	0.1&	0.1&	0.1&	 \\ \cline{2-10}
		& \multirow{2}{*}{GSHMC} &$L$&	3000&	2000&	2000&	2000&	2000&	2000&	\\
		&&$\varphi$	&0.1&	0.1&	0.1&	0.1&	0.1&	0.1& \\ \cline{2-9}
		&\multirow{3}{*}{MMHMC}& $h$ & 0.009&	0.012&	0.015&	0.018&	0.021&	0.024 \\
		& & $L$&2000&	1333&	1333&	1333&	1333&	1333 \\ 
		&&$\varphi$	&0.1&	0.1&	0.1&	0.1&	0.1&	0.1& \\	
		\bottomrule	
	\end{tabular}
	\label{Tab:GD_parameters}
\end{table*}

\begin{table*}[]
	\caption{Parameter values used for the Bayesian logistic regression model experiments. The Verlet integrator was employed on all methods. For HMC and MALA step size is drawn from $U(0.8 h,1.2 h)$. For HMC trajectory length is drawn from $U\{1,\dots,L \}$.}.
	\centering
	\begin{tabular}{c c r cccc}
		\toprule
		$D$ &  {Method}  & $h$ &0.02&	0.03&	0.04&	0.05   \\ 
		\midrule
		\multirow{4}{*}{German} & HMC &$L$& 25&25&	25&25 \\ \cline{2-7}
		& MALA &$L$& 1&	1 & 1& 1 \\ \cline{2-7}
		& \multirow{2}{*}{MMHMC}  &$L$& $U\{1,\dots,25\}$&$U\{1,\dots,25\}$&$U\{1,\dots,25\}$&$U\{1,\dots,25\}$ \\
		&& $\varphi$ & $U(0,0.5)$&	$U(0,0.5)$&	$U(0,0.9)$&	$U(0,0.9)$ \\
		\midrule
		&	& $h$ & 0.08&	0.1	&0.12&	0.14 \\
		\multirow{3}{*}{Sonar} & HMC &$L$& 200&	200&	200&200\\ \cline{2-7}
		& \multirow{2}{*}{MMHMC}  &$L$& 50&	50&	50&50 \\
		&& $\varphi$ & 0.25&	0.5&	0.5&	0.5 \\
		\midrule
		&	& $h$ & 0.05&	0.055&	0.06&	0.065 \\
		\multirow{3}{*}{Musk} & HMC &$L$&400&	400&	400&400 \\ \cline{2-7}
		& \multirow{2}{*}{MMHMC}  &$L$& 100&	100&	100&100 \\
		&& $\varphi$ & 0.25&	0.25&	0.25&	0.25 \\
		\midrule
		&	& $h$ & 0.01&	0.015&	0.02&	0.025 \\
		\multirow{3}{*}{Secom} & HMC &$L$& 900&	900&	900&900 \\ \cline{2-7}
		& \multirow{2}{*}{MMHMC}  &$L$& 150&	150&	150&	150 \\
		&& $\varphi$ & 0.25	&0.25	&0.25	&0.25 \\
		\bottomrule	
	\end{tabular}
	\label{Tab:BLR_parameters}
\end{table*}

\begin{table*}[]
	\caption{{\color{black}Parameter values used for the Stochastic Volatility model experiments. For HMC step size is drawn from $U(0.8 h,1.2 h)$ and trajectory length from $U\{1,\dots,L\}$. For MMHMC noise parameter is drawn from $U(0,\varphi)$. For all other cases, parameters are fixed.}}
	\centering
	\begin{tabular}{ccccccccc}
		\toprule
		$D$ &  {Method}  & Integrator & $h_{\boldsymbol\theta}$ & $h_{\mathbf x}$ & $L_{\boldsymbol\theta}$ & $L_{\mathbf x}$ & $\varphi_{\boldsymbol\theta}$ & $\varphi_{\mathbf x}$   \\ 
		\midrule
		\multirow{3}{*}{2003} & HMC & Verlet &0.01&	0.03&	6&	76 \\ 
		& {RMHMC}& Verlet & 0.5	&0.1&	6&	50  \\
		& {GSHMC}& Verlet & 0.008	&0.023&	3&	38 &0.25 & 0.4 \\
		& {MMHMC}& M-ME3  & 0.024&	0.069&	2&	25&	0.5&	0.8\\
		\midrule
		\multirow{2}{*}{5003} & HMC& Verlet & 0.006&	0.02&	6&	76\\ 
		& {MMHMC} & M-ME2 & 0.012&	0.032&	3&	38&	0.5&	0.5\\
		\midrule
		\multirow{2}{*}{10003} & HMC& Verlet & 0.004&	0.02&	6&	76\\
		& {MMHMC} & M-ME2 & 0.008&	0.022&	3&	38&	0.8&	0.8\\
		\bottomrule	
	\end{tabular}
	\label{Tab:SV_parameters}
\end{table*}

}

\printbibliography

\end{document}